\documentclass[12pt,tightenlines]{revtex4}

\usepackage{graphicx}
\usepackage{color} 
\usepackage{amsmath}
\usepackage{amssymb}
\usepackage{bbold}
\graphicspath{{graphics/}}

\def\Ket#1{\left|#1\right\rangle} 
\def\Bra#1{\left\langle#1\right|}
\def\KetBra#1#2{\Ket{#1}\!\Bra{#2}} 

\def\Proj#1{\KetBra{#1}{#1}}
\def\ProjInd#1#2{\Ket{#1}_{\!#2}\!\Bra{#1}}  
\def\Eins{\mathbf{1}} 

\def\ie{i.\,e.\ }

\def\bell#1{\ensuremath{\mathcal{B}_{#1}}}
\def\rhospace{\ensuremath{\mathcal{B}(\mathcal{H})}}

\newlength{\figurewidth}

\def\CCdots{\cdots}
\def\CLdots{\ldots}  
\def\BQuad{}

\begin{document}
\bibliographystyle{apsrev}

\title{A security proof of quantum cryptography based entirely on
  entanglement purification}

\author{Hans Aschauer}
\email{Hans.Aschauer@Physik.uni-muenchen.de}
\author{Hans J. Briegel}
\email{Hans.Briegel@Physik.uni-muenchen.de}

\affiliation{Sektion Physik, Ludwig-Maximilians-Universit\"at,
  Theresienstr.\ 37, D-80333 M\"unchen, Germany}

\date{\today}

\begin{abstract}
  We give a proof that entanglement purification, even with noisy
  apparatus, is sufficient to disentangle an eavesdropper (Eve) from
  the communication channel. In the security regime, the purification
  process factorises the overall initial state into a tensor-product
  state of Alice and Bob, on one side, and Eve on the other side, thus
  establishing a completely private, albeit noisy, quantum
  communication channel between Alice and Bob.  The security regime is
  found to coincide for all practical purposes with the purification
  regime of a two-way recurrence protocol.  This makes two-way
  entanglement purification protocols, which constitute an important
  element in the quantum repeater, an efficient tool for secure
  long-distance quantum cryptography.
\end{abstract}
\pacs{PACS: 3.67.Dd, 3.67.Hk, 3.65.Bz} 
\maketitle

\addtolength{\figurewidth}{0.65\linewidth}

\section{Introduction}
\label{sec:Intro}
A central problem of quantum
communication is how to faithfully transmit unknown quantum states
through a noisy quantum channel \cite{schumacher_noisy_channel}. While
information is sent through such a channel (for example an optical
fiber), the carriers of the information interact with the channel,
which gives rise to the phenomenon of decoherence and absorption; an
initially pure quantum state becomes a mixed state when it leaves the
channel. For quantum communication purposes, it is necessary
that the transmitted qubits retain their genuine quantum properties,
for example in form of an entanglement with qubits on the other side
of the channel.

In quantum cryptography \cite{bb84,ekert91}, noise in the
communication channel plays a crucial role: In the worst-case
scenario, all noise in the channel is attributed to an eavesdropper,
who manipulates the qubits in order to gain as much information on
their state as possible, while introducing only a moderate level of
noise \cite{ekert-huttner:94,fuchs-peres:96,luetkenhaus:96,fuchs-et-al:97}.

There are two well-established methods to deal with the problem of
noisy channels. The theory of quantum error correction \cite{CSS,steane95} has
mainly been developed to make quantum computation possible despite the
effects of decoherence and imperfect apparatus. Since data
transmission -- like data storage -- can be regarded as a special case
of a computational process, clearly quantum error correction can also
be used for quantum communication through noisy channels. An
alternative approach, which has been developed roughly in parallel
with the theory of quantum error correction, is the purification of
mixed entangled states \cite{bennett96,bennett96a,deutsch96}.

In quantum cryptography, noise in the communication channel plays a
crucial role: In the worst-case scenario, all noise in the channel
is attributed to an eavesdropper, who manipulates the qubits in order to
gain as much information on their state as possible, while introducing
only a moderate level of noise. 

To deal with this situation, two different techniques have been
developed: \emph{Classical privacy amplification} allows the
eavesdropper to have partial knowledge about the raw key built up
between the communicating parties Alice and Bob. From the raw key, a
shorter key is ``distilled'' about which Eve has vanishing (\ie
exponentially small in some chosen security parameter)
knowledge. Despite of the simple idea, proofs taking into account all
eavesdropping attacks allowed by the laws of quantum mechanics have
shown to be technically involved \cite{mayers,biham,hitoshi}.
Recently, Shor and Preskill \cite{shor} have given a simpler physical
proof relating the ideas in \cite{mayers,biham} to quantum error
correcting codes \cite{CSS} and, equivalently, to one-way entanglement
purification protocols \cite{bennett96a}. \emph{Quantum privacy amplification}
(QPA) \cite{deutsch96}, on the other hand, employs a two-way
entanglement purification recurrence protocol that eliminates any
entanglement with an eavesdropper by creating a few perfect EPR pairs
out of many imperfect (or impure) EPR pairs. The perfect EPR pairs can
then be used for secure key distribution in entanglement-based quantum
cryptography \cite{deutsch96,ekert91,mermin1992}. In principle, this
method guarantees security against any eavesdropping attack. However,
the problem is that the QPA protocol assumes ideal quantum operations.
In reality, these operations are themselves subject to noise. As shown
in \cite{briegel,duer_briegel,giedke}, there is an upper bound
$F_{\text{max}}$ for the achievable fidelity of EPR pairs which can be
distilled using noisy apparatus. \emph{A priori}, there is no way to
be sure that there is no residual entanglement with an eavesdropper.
This problem could be solved if Alice and Bob had fault tolerant
quantum computers at their disposal, which could then be used to
reduce the noise of the apparatus to any desired level. This was an
essential assumption in the security proof given by Lo and Chau
\cite{lo}.

In this paper, we show that the standard two-way entanglement
purification protocol alone, with some minor modifications to
accomodate certain security aspects as discussed below, can be used to
efficiently establish a \emph{perfectly private quantum channel}, even
when both the physical channel connecting the parties and the local
apparatus used by Alice and Bob are noisy.  %
\footnote{While it would be interesting to extend our proof to the
  hashing protocol, we note that for noisy local operations the
  hashing protocol, which requires Alice and Bob to apply a large
  number of CNOT operations in every distillation step, usually
  performs much worse that the recurrence protocols. The reason for
  this lies in the fact that the noise (\ie information loss)
  introduced with every CNOT operation accumulates and rapidly
  shatters the potential information that could ideally be gained from
  the parity measurement.}


In Section \ref{sec:QRandQPA} we will briefly review the concepts of
entanglement purification and of the quantum repeater, and discuss why
it is interesting to combine the security features of entanglement
purification with the long-distance feature of the quantum
repeater. Section \ref{sec:fact_of_eve} will give the main result of
our work: we prove that it is possible to \emph{factor out} an
eavesdropper using EPP, even when the apparatus used by Alice and Bob
is noisy. One important detail in the proof is the \emph{flag update
function}, which we will derive in Section
\ref{sec:flag_update_function}. We conclude the paper with a
discussion in Section \ref{sec:dicussion}.

\section{Entanglement purification and the quantum repeater}
\label{sec:QRandQPA}

\subsection{Entanglement purification}
\label{sec:ent_purification}
\begin{figure}[htbp]
  \begin{center}
    \includegraphics[width=\figurewidth]{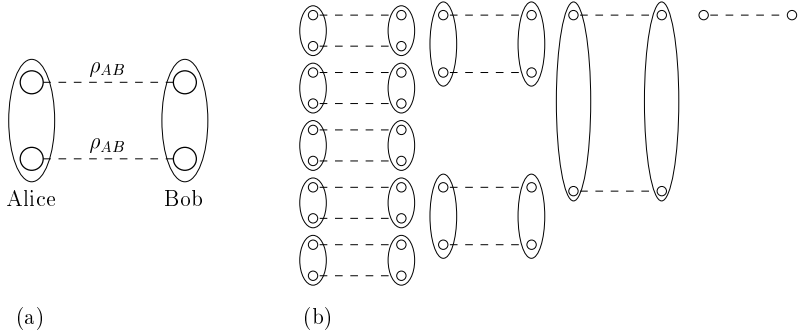}
    \caption{    
      \label{fig:ep_prot_and_proc}
      The entanglement purification protocol (a) and the
      entanglement distillation process (b).} 
  \end{center}
\end{figure}

As two-way entanglement purification protocols (2--EPP) play an
important role in this paper, we will briefly review one example of a
a recurrence protocol which was described in \cite{deutsch96}, and
called \emph{quantum privacy amplification} (QPA) by the authors. It
is important to note that we distinguish the entanglement purification
\emph{protocol} from the distillation \emph{process}: the first
consists of probabilistic local operations (unitary rotations and
measurements), where two pairs of qubits are combined, and either one
or zero pairs are kept, depending on the measurement outcomes. The
latter, on the other hand, is the procedure where the purification
protocol is applied to large ensemble of pairs recursively (see
Fig.~(\ref{fig:ep_prot_and_proc})).

In the quantum privacy amplification 2--EPP, two pairs of qubits,
shared by Alice and Bob, are considered to be in the state
\(\rho_{A_1B_1} \otimes \rho_{A_2B_2}\). Without loss of generality
(see later), we may assume that the state of the pairs is of the
Bell-diagonal form,
\begin{equation}
  \label{eq:bell_diagonal}
  \rho_{AB} = A \Proj{\Phi^+} + B \Proj{\Psi^-} + C \Proj{\Psi^+} + D
  \Proj{\Phi^-}.
\end{equation}
Following \cite{deutsch96}, the protocol consists of three steps: 
\begin{enumerate}
\item Alice applies to her qubits a \(\pi/2\) rotation, $U_x$, Bob a
  \(-\pi/2\) rotation about the \(x\) axis, $U_x^{-1}$.
\item Alice and Bob perform the bi-lateral CNOT operation
  \[\mathrm{BCNOT_{A_1B_1}^{A_2B_2}} = \mathrm{CNOT}_{A_1}^{A_2}
  \otimes  \mathrm{CNOT}_{B_1}^{B_2} \]
  on the four qubits.
\item Alice and Bob measure both qubits of the target pair \(A_2B_2\)
  of the BCNOT operation in the $z$ direction. If the measurement
  results coincide, the source pair \(A_1B_1\) is kept, otherwise it
  is discarded. The target pair is always discarded, as it is
  projected onto a product state by the bilateral measurement.
\end{enumerate}

By a straigtforward calculation, one gets the result that the state
of the remaining pair is still a Bell diagonal state, with the
diagonal coefficients \cite{deutsch96}
\begin{equation}
  \label{eq:qpa_recursion}
  \begin{split}
    A' &= \frac{A^2 + B^2}{N},\quad  B' = \frac{2CD}{N}\\ 
    C' &= \frac{C^2 + D^2}{N},\quad  D' = \frac{2AB}{N},\\
  \end{split}
\end{equation}
and the normalization coefficient \(N= (A+B)^2 + (C+D)^2\), which is
the probability that Alice's and Bob's measurement results in step 3
coincide. Note that, up to the normalization, these recurrence
relations are a quadratic form in the coefficients $A, B, C, $ and
$D$. These relations allow for the following interpretation (which can
be used to obtain the relations (\ref{eq:qpa_recursion}) in the
first place): As all pairs are in the Bell diagonal state
(\ref{eq:bell_diagonal}), one can interpret \(A, B, C,\) and \(D\) as
the relative frequencies with which the states
\(\Ket{\Phi^+}, \Ket{\Psi^-}, \Ket{\Psi^+},\) and \(\Ket{\Phi^-}\),
respectively, appear in the ensemble. By looking at (\ref{eq:qpa_recursion}) one finds that
the result of combining two $\Ket{\Phi^+}$ or two $\Ket{\Psi^-}$ pairs
is a $\Ket{\Phi^+}$ pair, combining a $\Ket{\Psi^+}$ and a
$\Ket{\Phi^-}$ (or vice versa) yields a $\Ket{\Psi^-}$ pair, and so
on. Combinations of $A, B, C,$ and $D$ that do not occur in
(\ref{eq:qpa_recursion}), namely $AC$, $AD$, $BC$ and $BD$, are
``filtered out'', \ie they give different measurement results for the
bilateral measurement in step 3 of the protocol. We will use this way of
calculating recurrence relations for more complicated situations
later.

Numerical calculations \cite{deutsch96} and, later, an analytical
investigation \cite{macciavello98} have shown that for all initial
states (\ref{eq:bell_diagonal}) with \(A > 1/2\), the recurrence
relations (\ref{eq:qpa_recursion}) approach the fixpoint \(A=1, B = C
= D = 0\); this means that given a sufficiently large number of initial
pairs, Alice and Bob can distill asymptotically pure EPR  pairs.

\subsection{Entanglement purification with noisy apparatus and the
  quantum repeater}
\label{sec:quantum_repeater}

Under realistic conditions, the local operations (quantum gates, measurements)
themselves, that constitute a purification protocol, will never be perfect 
and thus introduce a certain amount of noise to the ensemble when they 
are applied by Alice and Bob. The follow questions then arise: How does a 
protocol perform under the influence of local noise? How robust is it and
what is the threshold for purification? These questions have been dealt with
in Refs.~\cite{briegel,duer_briegel,giedke}. The main results are, in brief,
that for a finite level of local noise, there is a maximum achievable
fidelity $F_{\text{max}}<1$ beyond which purification is not possible. 
Similarly, the minimum required fidelity $F_{\text{min}}>1/2$ for 
purification has increased with respect to the ideal protocol \cite{bennett96}. 
The purification regime $[F_{\text{min}}, F_{\text{max}}] \in [1/2,1]$ has 
thus become smaller compared to the noiseless case. With an increasing noise 
level, the size of the purification regime shrinks until, at the purification 
threshold, $F_{\text{min}}$ and $F_{\text{max}}$ coincide and the protocol 
breaks down. At this point, the noise of the local operations corresponds to 
a loss of information that is larger than the gain of information ontained 
by a destillation step in the ideal case. 

For a moderate noise level (of the order of a few percent for the 
recurrence protocols of Refs.~\cite{bennett96,deutsch96}), entanglement 
purification remains an efficient tool for establishing high-fidelity 
(although not perfect) EPR pairs, and thus for quantum communication over 
distances of the order of coherence length of a noisy channel. The 
restriction to the coherence length is due to the fact that the fidelity 
of the initial ensemble needs to be above the value $F_{\text{min}}(>1/2)$. 

\emph{Long-distance quantum communication} describes a situation where the
length of the channel connecting the parties is typically much longer
than its coherence- and absorption length. As the depolarisation errors 
and the absorption losses scale exponentially with the length of the 
channel, one cannot send qubits directly through the channel.

To solve this problem, there are two solutions known. The first is to
treat quantum communication as a (very simplistic) special case of
quantum computation. The methods of fault tolerant quantum computation
\cite{preskill,knill} and quantum error correction could then be used
for the communication task. An explicit scheme for data transmission
and storage has been discussed by Knill and Laflamme \cite{knill1996},
using the method of concatenated quantum coding. While this idea shows
that it is \emph{in principle} possible to get polynomial or even
polylogarithmic \cite{kitaev91,aharonov96,knill1998}
scaling in quantum communication, it has an important
drawback: long-distance quantum communication using this idea is as
difficult as fault tolerant quantum computation, despite the fact that
\emph{short} distance QC is (from a technological point of view)
already ready for practical use. 


The other solution for the long-distance problem is the entanglement
based quantum repeater (QR) \cite{briegel,duer_briegel} with two-way
classical communication. It employs both entanglement purification
\cite{bennett96,bennett96a,deutsch96} and entanglement swapping
\cite{bennett93,zukowski1993,Pan1998} in a meta-protocol, the nested
two-way entanglement purification protocol (NEPP). The apparatus used 
for quantum operations in the NEPP tolerates noise on the (sub-) percent 
level. As this tolerance is two orders of magnitude less 
restrictive than for fault tolerant quantum computation, it seems to make 
the quantum repeater a promising concept also for practical realisation in
the future. Please note that the quantum repeater has been designed not 
only to solve the problem of \emph{decoherence}, but also of \emph{absorption}. 
For the latter, the possiblity of quantum storage is required at the repeater 
stations. An explicit implementation that takes into account absorption is 
given by the photonic channel of Ref.~\cite{vanEnk1997,vanEnk1998} 
(see also \cite{briegel2000}).

\subsection{The quantum repeater and quantum privacy amplification}  
\label{sec:nEPP_and_QPA}

The aim of this paper, as mentioned in the introduction, is to show that 
entanglement distillation using realistic apparatus is sufficient to create 
\emph{private entanglement} 
\footnote{We owe the term \emph{private entanglement} to Charles Bennett.} 
between Alice and Bob, \ie pairs of entangled qubits of which Eve is 
guaranteed to be disentangled even though they are not pure EPR pairs. 
If these pairs are used to teleport quantum information from Alice to Bob, 
they can be regarded as a \emph{noisy but private quantum channel}.

This will also prove the security of quantum communication using the
entanglement-based quantum repeater, since it is only necessary to
consider the outermost entanglement purification step in the NEPP,
which is performed by Alice and Bob exclusively, i.e.\ without
the support of the parties at the intermediate repeater stations. 
In particular, it is not
necessary to analyze the effect of noisy Bell measurements on the
security. In the worst case scenario, Alice and Bob assume that all
repeater stations are completely under Eve's control, anyway. For this
reason, Alice and Bob are not allowed to make assumptions on the
method how the pairs have been distributed.

The role of the quantum repeater in the \emph{security} proof 
is thus the following. It tells us that it is possible
to distribute EPR pairs of high fidelity over arbitrary distances (with
polynomial overhead), given that the noise level of the apparatus (or
operations) used in the entanglement purification is below a certain
(sub percent) level. The noise in the apparatus will reflect itself in
the fact that the final distributed pairs between Alice and Bob are
also imperfect (i.e.\ not pure Bell states), and so the question arises 
whether these imperfect pairs can be used for e.g.\ \emph{secure} key 
distribution. The answer is yes. Since the security regime practically 
coincides with the purification regime  (see Sec.~\ref{sec:results_full}), 
quantum communication is guaranteed to be secure whenever the noise level 
of the local operations is in the operation regime of the quantum repeater.

\section{Factorization of Eve}
\label{sec:fact_of_eve}
In this section we will show that 2--EPP with noisy apparatus is
sufficient to factor out Eve in the Hilbertspace of Alice, Bob, their
laboratories, and Eve. For the proof, we will first introduce the concept of
the lab demon as a simple model of noise. Then we will consider the
special case of binary pairs, where we have obtained analytical
results. Using the same techniques, we generalize the result to the
case of Bell-diagonal ensembles. To conclude the proof, we show how
the most general case of ensembles, described by an arbitray entangled 
state of all the qubits on Alice's and Bob's side, can be reduced to the 
case of Bell-diagonal ensembles.

\subsection{The effect of noise}
\label{sec:eff_of_noise}
In this section we will answer the following question: what is the
effect of an error, introduced by some noisy operation at a given
point of the distillation process? We restrict our attention to the 
following type of noise:
\begin{itemize}
\item It acts locally, \ie the noise does not introduce correlations
  between remote quantum systems.
\item It is memoryless, \ie on a timescale imposed by the sequence
  of steps in a given protocol, there are no correlations between the
  ``errors'' that occur at different times.
\end{itemize}

The action of noisy apparatus on a quantum system in state \(\rho \in
\rhospace\) can be formally described by some trace conserving,
completely positive map. Any such map can be written in the
operator-sum representation
\cite{kraus_states,schumacher_noisy_channel},
\begin{equation}
  \label{eq:operator_sum}
  \rho \rightarrow \sum_i A_i \rho A_i^\dagger,
\end{equation}
with linear operators \(A_i\), that fulfill the normalization
condition \(\sum_i A A^\dagger = \Eins\). The operators \(A_i\) are
the so-called \emph{Kraus operators} \cite{kraus_states}.

As we have seen above, in the purification protocol the CNOT
operation, which acts on two qubits $a$ and $b$, plays an important
role. For that reason, it is necessary to consider noise which acts on
a two-qubit Hilbert space \(\mathcal{H} = \mathbb{C}^2_a \otimes
\mathbb{C}^2_b\). Eq.~(\ref{eq:operator_sum}) describes the most general
non-selective operation that can, in principle, be implemented. For
technical reasons, however, we restrict our attention to the case that
the Kraus operators are proportional to products of Pauli
matrices. The reason for this choice is that Pauli operators map Bell
states onto Bell states, which will allow us to introduce the very
useful concept of \emph{error flags}
later. Eq.~(\ref{eq:operator_sum}) can then be written as
\begin{equation}
\rho_{ab} \rightarrow  \sum_{\mu,\nu=0}^{3}
f_{\mu\nu}\sigma_{\mu}^{(a)}\sigma_{\nu}^{(b)} \rho_{ab}\sigma_{\mu}^{(a)}
\sigma_{\nu}^{(b)}\,,
\label{eq:noise_model}
\end{equation}
with the normalization condition \(\sum_{\mu,\nu=0}^{3} f_{\mu\nu} =
1\). Note that Eq.~(\ref{eq:noise_model}) includes, for an appropriate
choice of the coefficients $f_{\mu\nu}$, the one- and two-qubit
depolarizing channel and combinations thereof, as studied in
\cite{briegel,duer_briegel}; but it is more general. Below, we will
refer to these special Kraus operators as \emph{error operators}.

The proof can be extended to more general noise models if a slightly
modified protocol is used, where the twirl operation of step 1 is
repeated after every distillation round \footnote{We are grateful to
  C.~H.~Bennett for pointing out this possibility.}. The concatenated
operation, which consists of a general noisy operation followed by
this \emph{regularization} operation, is Bell diagonal \ie it maps
Bell-diagonal states onto Bell-diagonal states, but since it maps
\emph{all} states to a Bell-diagonal state, it clearly cannot be
written in the form (\ref{eq:noise_model}). However, for the purpose
of the proof, it is in fact only necessary that the concatenated map
restricted to the space of all Bell-diagonal ensembles can be written
in the form (4); we call such a map a \emph{restricted Bell-diagonal
  map}. Clearly, not all restricted Bell-diagonal maps are of the form
(\ref{eq:noise_model}), which can be seen considering a map which maps
any Bell diagonal state to a pure Bell state. Such a map could,
however, not be implemented locally. Thus the question remains whether
a restricted Bell-diagonal map which can be implemented locally can be
written in the form (\ref{eq:noise_model}). Though we are not aware of
a formal proof of such a theorem, we conjecture that it holds true:
the reduced density operator of each qubit must remain in the
maximally mixed state, which is indeed guaranteed by a mixture of
unitary rotations. We also have numerical evidence which supports this
conjecture. Note that in the case of
such an active regularization procedure, it is important that the
Pauli rotations in step~1 can be performend well enough to keep the
evolution Bell diagonal. This is, however, not a problem, since Alice
and Bob are able to propagate the Pauli rotations through the unitary
operations of the EPP, which allows them to perform the rotations just
before a measurement, or, equivalently, to rotate the measurement
basis. This is similar to the concept of error correctors (where the
error consists in \emph{ommiting} a required Pauli operaton), as
described in Section \ref{sec:unitary_and_errors}, and to the
by-product matrix formalism developed in \cite{raussendorf2001}.

The coefficients $f_{\mu\nu}$ in (\ref{eq:noise_model}) can be
interpreted as the joint probability that the Pauli rotations
$\sigma_\mu$ and $\sigma_\nu$ occur on qubits $a$ and $b$,
respectively. For pedagogic purposes we employ the following
interpretation of (\ref{eq:noise_model}): Imagine that there is a
(ficticious) little demon in Alice's laboratory -- the ``lab demon''
-- which applies in each step of the distillation process randomly,
according to the probability distribution \(f_{\mu\nu}\), the Pauli
rotation \(\sigma_\mu\) and \(\sigma_\nu\) to the qubits $a$ and $b$,
respectively. The lab demon summarizes all relevant aspects of the lab
degrees of freedom involved in the noise process.

Noise in Bob's laboratory, can, as long as we restrict ourselves to
Bell diagonal ensembles, be attributed to noise introduced by Alice's
lab demon, without loss of generality; this is, however, not a crucial
restriction, as we will show in Section \ref{sec:non_bell_diag}. It is
also possible to think of a second lab demon in Bob's lab who acts similarly
to Alice's lab demon. This would, however, not affect the arguments
employed in this paper.

The lab demon does not only apply rotations randomly, he also
maintains a list in which he keeps track of which rotation he has
applied to which qubit pair in which step of the distillation process.
What we will show in the following section is that, from the mere
content of this list, the lab demon will be able to extract -- in the
asymptotic limit -- full information about the state of each residual
pair of the ensemble. This will then imply that, given the lab demons
knowledge, the state of the distilled ensemble is a tensor product of
pure Bell states.  Furthermore, Eve cannot have information on the
specific sequence of Bell pairs (beyond their relative frequencies)
--- otherwise she would also be able to learn, to some extent, at
which stage the lab demon has applied which rotation.

From that it follows that Eve is \emph{factored out}, \ie the overall state
of Alice's, Bob's and Eve's particles is described a density operator
of the form
\begin{equation}
  \label{eq:rho_ABEL}
  \rho_\mathrm{ABE} = 
      \left( 
        \sum_{i,j = 0}^1 f^{(i,j)}
        \ProjInd{\bell{i,j}}{\mathrm{AB}} 
      \right) 
  \otimes \rho_\mathrm{E}\,,
\end{equation}
where $\sum_{i,j}f^{(i,j)}=1$, and \(\bell{i,j}\) describe the four
Bell states as defined in Sec.~\ref{sec:non_bell_diag}.

Note that the lab demon was only introduced for pedagogical reasons.
In reality, there will be other mechanisms of noise. However, all
physical processes that result in the same completely positive map
(\ref{eq:noise_model}) are equivalent, \ie cannot be distinguished
from each other if we only know how they map an input state \(\rho_i\)
onto an output state \(\rho_f\). In particular, the processes must lead
to the same level of security (regardless whether or not error flags
are measured or calculated by anybody): otherwise they would be
distinguishable.

In order to separate conceptual from technical considerations and to
obtain analytical results, we will first concentrate on the special
case of binary pairs and a simplified error model. After that, we
generalize the results to \emph{any} initial state.

\subsection{Binary pairs}
\label{sec:binary_pairs}
In this section we restrict our attention to pairs in the state 
\begin{equation}
  \label{eq:binary_pairs}
  \rho_{AB} = A \ProjInd{\Phi^+}{AB} + B \ProjInd{\Psi^-}{AB},
\end{equation}
and to errors of the form
\label{sec:any_initial_state}
\begin{equation}
  \label{eq:binary_noise}
  \rho_{AB}^{(1)} \otimes \rho_{AB}^{(2)} \rightarrow
  \sum_{\mu,\nu \in \{0,1\}}f_{\mu\nu}U^{(1)}_\mu U^{(2)}_\nu \rho_{AB}^{(1)}
  \otimes \rho_{AB}^{(2)} {U^{(1)}_\mu}^\dagger {U^{(2)}_\nu}^\dagger
\end{equation}
with \(U_0^{(1,2)} = \text{id}^{(1,2)}\) and \(U_1^{(1,2)} =
{\sigma_x}^{(1,2)}\). Eq.~(\ref{eq:binary_noise}) describes a
\emph{two-bit correlated spin-flip channel}. The indices 1 and 2
indicate the source and target bit of the bilateral CNOT (BCNOT)
operation, respectively. It is straightforward to show that, using
this error model in the 2--EPP, binary pairs will be mapped onto
binary pairs.

At the beginning of the distillation process, Alice and Bob share an
ensemble of pairs described by (\ref{eq:binary_pairs}). Let us imagine
that the lab demon attaches one classical bit to each pair, which he
will use for book-keeping purposes. At this stage, all of these bits,
which we call ``error flags'', are set to zero. This reflects the
fact that the lab demon has the same \emph{a priori} knowledge about
the state of the ensemble as Alice and Bob.

In each purification step, two of the pairs are combined. The lab
demon first simulates the noise channel (\ref{eq:binary_noise}) on
each pair of pairs by the process described. Whenever he applies a
\(\sigma_x\) operation to a qubit, he inverts the error flag of the
corresponding pair. Alice and Bob then apply the 2--EPP to each pair
of pairs; if the measurement results in the last step of the protocol
coincide, the source pair will be kept. Obviously, the error flag of
that remaining pair will also depend on the error flag of the the
target pair, \ie the error flag of the remaining pair is a function of
the error flags of both ``parent'' pairs, which we call the \emph{flag
  update function}. In the case of binary pairs, the flag update
function maps two bits (the error flags of \emph{both} parents) onto
one bit. In total, there exist 16 different functions $f\!:\!\{0,1\}^2
\to \{0,1\}$. From these, the lab demon chooses the logical AND
function as the flag update function, \ie the error flag of the
remaining pair is set to ``1'' if and only if both parent's error
flags had the value ``1''.

After each purification step, the lab demon  divides all pairs into
two subensembles, according to the value of their error flags. By a 
straightforward calculation, we obtain for the coefficients $A_i$ and
$B_i$, which completely describe the state of the pairs in the
subensemble with error flag $i$, the following recurrence relations:

\begin{equation}
  \label{eq:recursion_binary}
  \begin{split}
    A_0' =  & \frac{1}{N}
    (f_{00}(A_0^2 + 2A_0A_1) + f_{11}(B_1^2+2B_0B_1)\\
    & +f_s(A_0B_1+A_1B_1+A_0B_0))\\
    A_1' =  & \frac{1}{N}
    \left(f_{00}A_1^2 + f_{11}B_0^2 + f_s A_1B_0 \right)\\
    B_0' =   &\frac{1}{N}
    (f_{00}(B_0^2 + 2B_0B_1) + f_{11}(A_1^2+2A_0A_1)\\
    &+f_s(B_0A_1+B_1A_1+B_0A_0))\\
    B_1' =  & \frac{1}{N}
    \left(f_{00}B_1^2 + f_{11}A_0^2 + f_s B_1A_0 \right)
  \end{split}
\end{equation}
with  \(N = (f_{00}+f_{11})((A_0+A_1)^2+(B_0+B_1)^2) + 2f_s (A_0+A_1)
(B_0+B_1) \) and \(f_s = f_{01} + f_{10}\). 

For the case of uncorrelated noise, \(f_{\mu\nu} = f_\mu f_\nu\), we
obtain the following analytical expression for the relevant fixpoint 
of the map (\ref{eq:recursion_binary}):
\begin{equation}
  \label{eq:binary_pairs_fixpoint}
  \begin{split}
    A_0^\infty & = \frac
    {4f_0^2 - 4f_0 + (2f_0 - 1)\sqrt{4f_0-3}+1}
    {2 (2 f_0 - 1)^2}, \\
    A_1^\infty & = 0,\quad
    B_0^\infty   = 0 ,\quad
    B_1^\infty   = 1 - A_0^\infty  .  
  \end{split}
\end{equation}

\newcommand{\pd}[2]{\frac{\partial #1}{\partial #2}}

Note that, while Eq. (\ref{eq:binary_pairs_fixpoint}) gives a fixpoint
of (\ref{eq:recursion_binary}) for \(f_0 \ge 3/4\), this does not
imply that this fixpoint is an attractor. In order to investigate the
attractor properties, we calculate the eigenvalues of the matrix of
first derivatives,
\begin{equation}
  \label{eq:derive_binary}
  M_D = \left. \left(
      \begin{array}{c c c}
        \pd{A_0'}{A_0} &\cdots &\pd{B_1'}{A_0}\\ 
        \vdots & \ddots & \vdots \\
        \pd{A_0'}{B_1} &\cdots &\pd{B_1'}{B_1}\\ 
      \end{array}
      \right) \right|_\mathrm{fixpoint} \quad .
\end{equation}

We find that the modulus of the eigenvalues of this matrix is smaller
than unity for \(f_0^\mathrm{crit}=0.77184451 < f_0 \le 1\), which
means that in this interval, the fixpoint
(\ref{eq:binary_pairs_fixpoint}) is also an attractor. This is in
excellent agreement with a numerical evaluation of
(\ref{eq:recursion_binary}), where we found that \(0.77182 <
f_0^\mathrm{crit} < 0.77188\).

\begin{figure}[htbp]
  \begin{center}
    \includegraphics{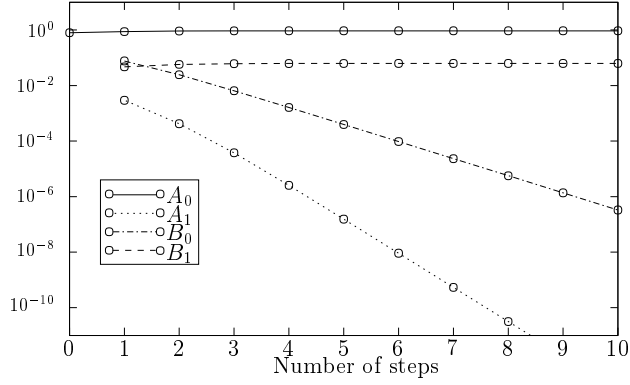}
    \caption{ \label{fig:binary_data}
      The evolution of the four parameters $A_0, A_1, B_0$, and $B_1$
      in the security regime. Note that both $A_1$ and $B_0$ decrease
      exponentially fast in the number of steps. The initial fidelity
      was 80\%, and the values of the noise parameters were
      \(f_{00}=0.8575\), \(f_{01}=f_{10}=f_{11} = 0.0475\).}
  \end{center}
\end{figure}

We have also evaluated (\ref{eq:recursion_binary}) numerically in order to
investigate correlated noise (see Fig.~\ref{fig:binary_data}). Like in
the case of uncorrelated noise, we found that the coefficients $A_0$
and $B_1$ reach, during the distillation process, some finite value,
while the coefficients $A_1$ and $B_0$ decrease exponentially fast,
whenever the noise level is moderate.

In other words, both subensembles, characterized by the value of the
respective error flags, approach a pure state asymptotically: The
pairs in the ensemble with error flag ``0'' are in the state
\(\Ket{\Phi^+}\), while those in the ensemble with error flag ``1''
are in the state \(\Ket{\Psi^+}\).

\subsubsection*{A map of the fixpoints}

\begin{figure}[htbp]
  \begin{center}
    \includegraphics*[width=0.9\figurewidth]{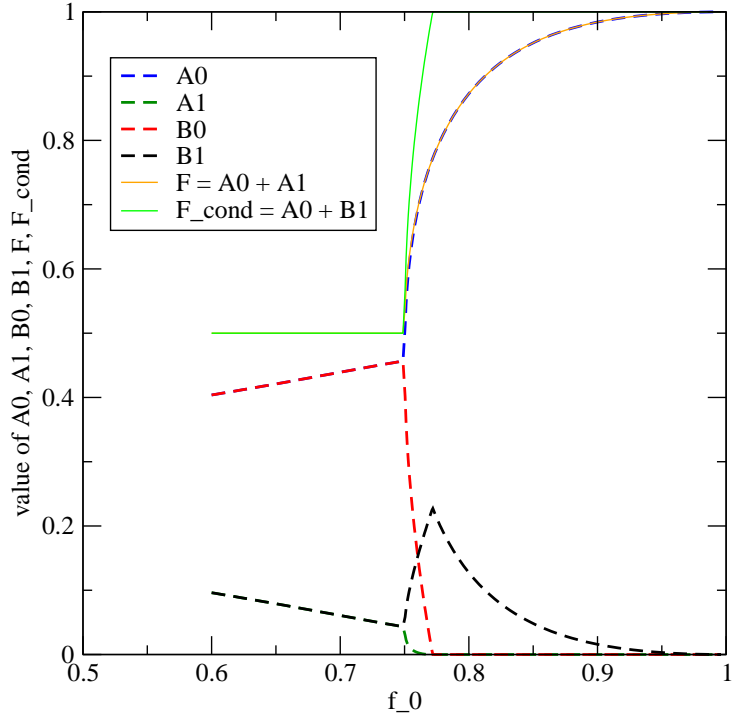}
    \caption{    \label{fig:fixpoint_map}
      The values of \(A_0,A_1,B_0,B_1,F,F^\mathrm{cond}\) at the
      fixpoint as a function of the noise parameter \(f_0\).}
  \end{center}
\end{figure}

In Fig.~\ref{fig:fixpoint_map}, the values of \(A_0^\infty,
A_1^\infty, B_0^\infty, B_1^\infty, F^\infty,\) and
\(F^\mathrm{cond,\infty}\) have been plotted as a function of the
noise parameter \(f_0\). Most interesting in this graph is the shape
of the curve representing the conditional fidelity: For all noise
parameters \(f_0 \le 0.75\), the conditional fidelity reaches at the
fixpoint the value \(0.5\), while for noise parameters \(f_0 \ge
0.77184451\), the conditional fidelity reaches unity. In the
intermediate regime (\(0.75 < f_0 < 77184451\)), the curve can be
fitted by a square root function \(F^\mathrm{cond}(f_0) = 0.5 + 3.4
\sqrt{f_0-0.75}.\) 

The emergence of the intermediate regime of noise parameters, where
the 2--EPP is able to purify and the lab demon does not gain full
information on the state of the pairs is somewhat surprising and shows
that the factorization of the eavesdropper is by no means a
trivial consequence of (noisy) EPP. From a mathematical point
of view, it is consistent with the finding after
Eq.~(\ref{eq:derive_binary}).

\subsubsection*{The purification curve}
\label{sec:purification_curve}

\begin{figure}[htbp]
  \begin{center}
    \includegraphics[width=0.9\figurewidth]{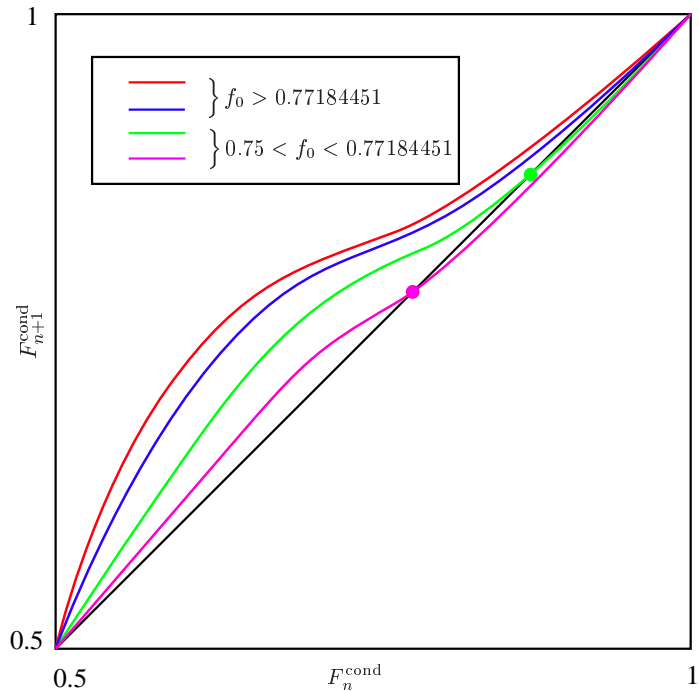}
    \caption{    \label{fig:illustration_puri_curve}
      Illustration of the purification curve for variouse noise
      levels $f_0$. In order to make the point clear, the effect has
      been strongly overdrawn. See text.} 
  \end{center}
\end{figure}

To understand the emergence of the intermediate regime better, we have 
plotted the purification curve for binary pairs, \ie the 
\(F^\mathrm{cond}_n - F^\mathrm{cond}_{n+1}\)-diagram. A problem with this 
diagram is that
the state of the pairs is specified by three independent parameters
(\(A_0, A_1, B_0, B_1\) minus normalization), so that such plots can
only show a specific section through the full parameter space. Below
we explain in detail how these sections have been
constructed. Fig. \ref{fig:illustration_puri_curve} shows an overdrawn
illustration of what we found: for noise parameters close to the
purification threshold, the purification curves have a point of
inflection. If the noise level increases (\ie \(f_0\) decreases), the
curves are quasi ``pulled down''. For \(f_0 = 0.77184451\), the slope
of the purification curve at the fixpoint \(F^\mathrm{cond} = 1\)
equals unity. If we further decrease \(f_0\), the fixpoint is no
longer an attractor, but due to the existence of the point of
inflection, a new attractive fixpoint appeares.

\begin{figure}[htbp]
  \begin{center}
    \vspace{-1cm}
    \includegraphics[width=0.9\figurewidth]{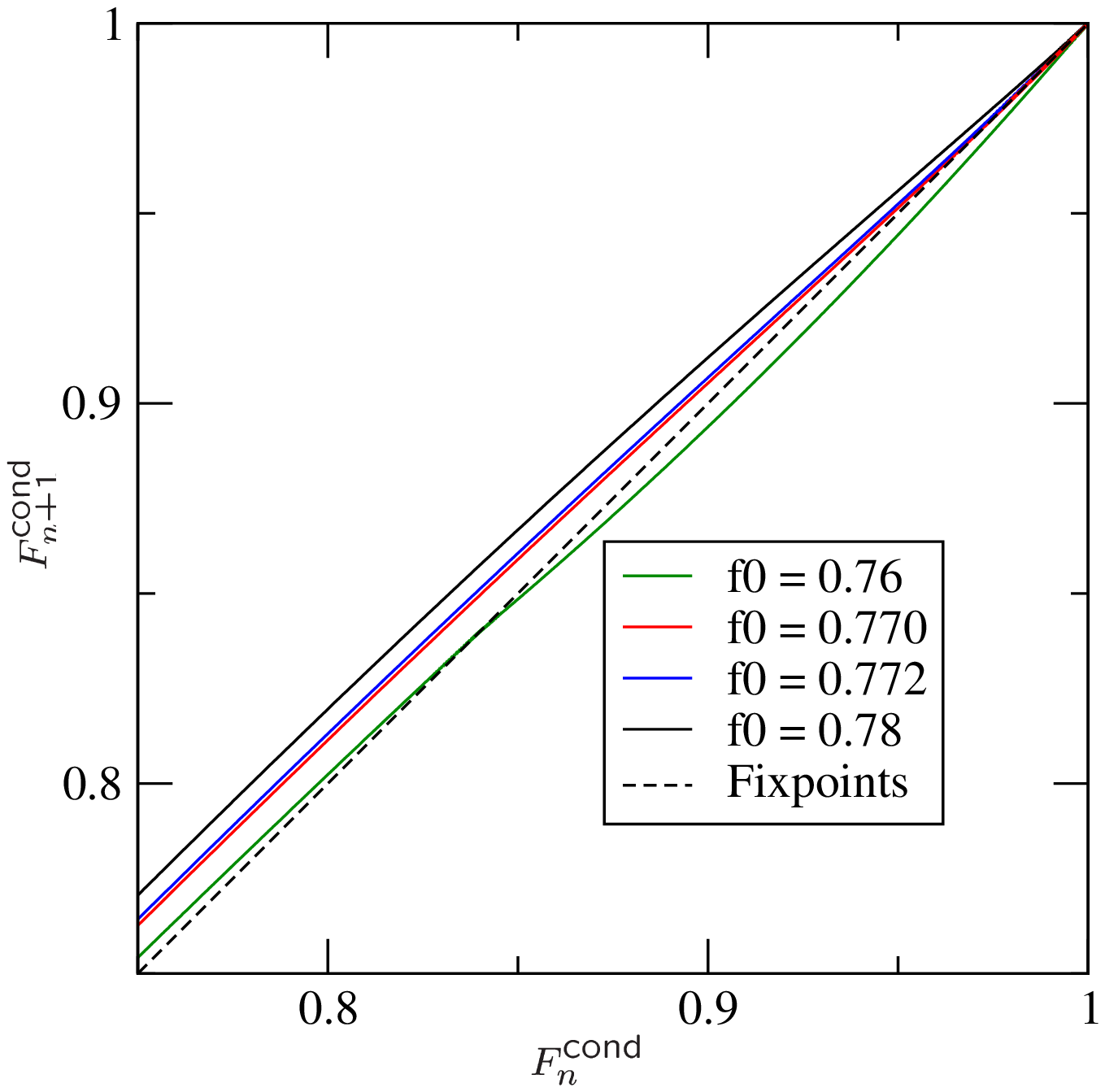}

    \vspace{1.5cm}
    \includegraphics[width=0.9\figurewidth]{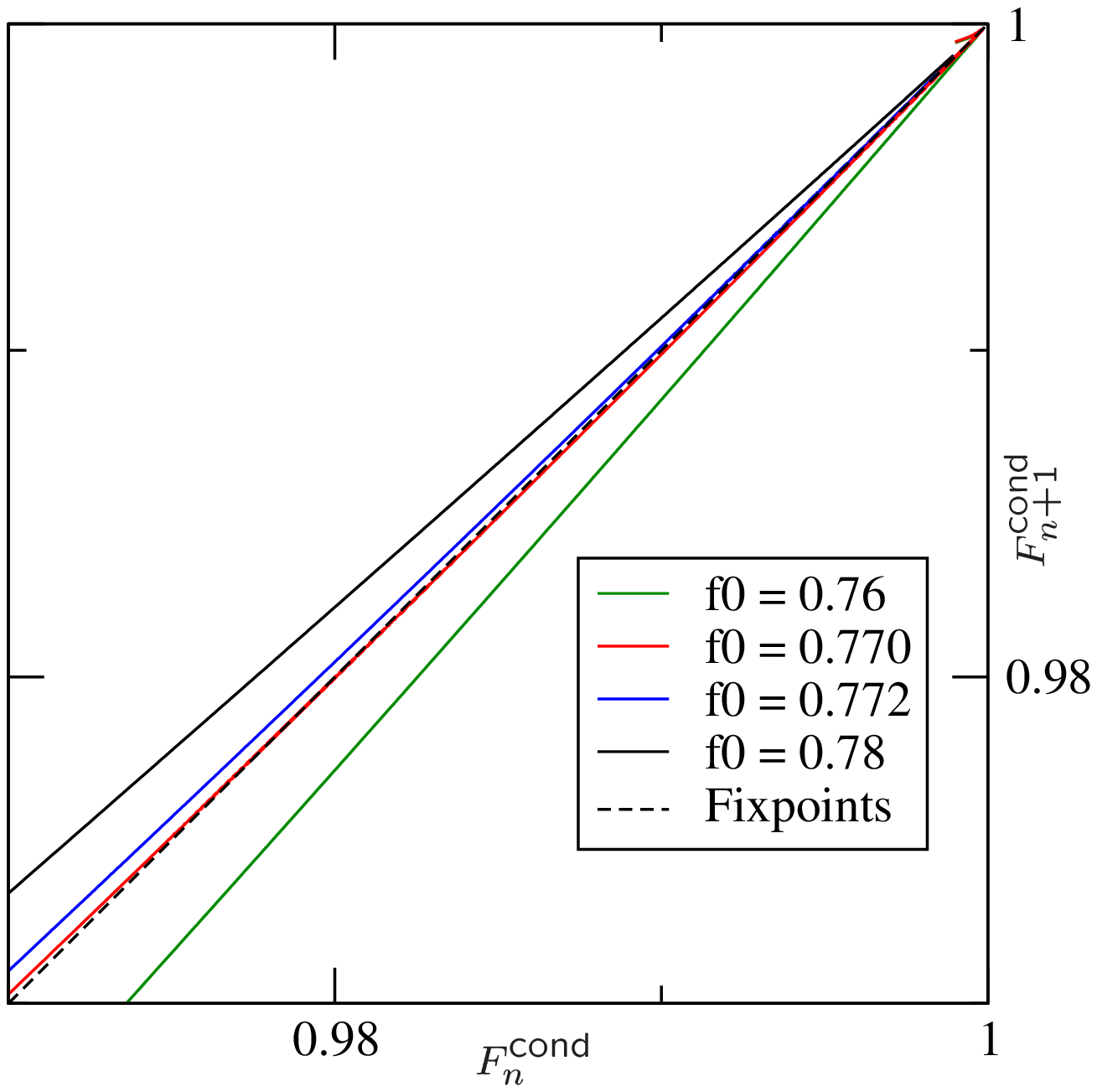}
    \vspace{5ex}
    \caption{ \label{fig:puri_curve_actual}
      Actual data from which
      Fig.~(\ref{fig:illustration_puri_curve}) has been infered.} 
  \end{center}
\end{figure}

To obtain the one-parametric curves shown in
Fig~\ref{fig:puri_curve_actual}, we used the following technique:
starting with the point \((A_0,A_1,B_0, B_1)^{n=0} = (0.6,0,0.4,0)\),
we calculated \((A_0,A_1,B_0, B_1)^{n=1}\) by applying the recursion
relations (\ref{eq:recursion_binary}) once. The points on the straight
line in parameter space connecting these two points have then been used 
as input values for the map given by the \(n\)th power of
(\ref{eq:recursion_binary}). For the plot, the resulting curve
segments have been concatenated. This procedure has been repeated for
all noise parameters \(f_0\) that are specified in
Fig.~\ref{fig:puri_curve_actual}. Note that at the critical value
\(f_0^\mathrm{crit}=0.77184451\), the number of iterations required to
reach any \(\epsilon\)-environment of the fixpoint
\emph{diverges}. This fact will later be discussed in a more general
case, see Fig.~\ref{fig:f_f_cond_iter}.

To conclude this section, we summarize: For all values  
\(0.77184451 \equiv f_0^\mathrm{crit} \le f_0 \le 1\) the 2--EPP
purifies and at the same time any eavesdropper is factored out.  
In a small interval, \(0.75 < f_0 < f_0^\mathrm{crit}\equiv 0.77184451\),
just above the threshold of the purification protocol, the conditional
fidelity does not reach unity, while the protocol is in the
purification regime. Even though this interval is small and of little
practical relevance (for these values of $f_0$ we are already out of the 
repeater regime \cite{briegel} and purification is very inefficient), its
existence shows that the process of factorization is not trivially connected 
to the process of purification.

\subsection{Bell-diagonal initial states}
Now we want to show that the same result is true for arbitrary Bell
diagonal states (Eq.~(\ref{eq:bell_diagonal})) and for noise of the
form (\ref{eq:noise_model}). The procedure is the same as in the case
of binary pairs; however, a few modifications are required.

In order to keep track of the four different error operators
\(\sigma_\mu\) in (\ref{eq:noise_model}), the lab demon has to attach
two classical bits to each pair; let us call them the phase error bit
and amplitude error bit. Whenever a $\sigma_x$ ($\sigma_z$,
$\sigma_y$) error occurs, the lab demon inverts the error amplitude
bit (error phase bit, both error bits). To update these error flags,
he uses the update function given in Tab.~\ref{tab:error_flags}. The
physical reason for the choice of the flag update function will be
given in the next section.

\begin{table}[h]
  \begin{tabular}[t]{|r|cccc|}
    \toprule 
    &(00)&(01)&(10)&(11) \\ 
    \colrule 
    (00)&(00)&(00)&(00)&(10) \\ 
    (01)&(00)&(01)&(11)&(00) \\ 
    (10)&(00)&(11)&(01)&(00) \\
    (11)&(10)&(00)&(00)&(00) \\ 
    \botrule
  \end{tabular}
  \caption
  {The value (phase error,amplitude error) of the updated error flag
  of a pair that is kept after a 2--EPP step, given as a function of the
  error flags of $P_1$ and $P_2$ (left to right and top to bottom,
  respectively). }
  \label{tab:error_flags}
\end{table}

Here, the lab demon divides all pairs into four subensembles,
according to the value of their error flag. In each of the
subensembles the pairs are described by a Bell diagonal density
operator, like in Eq.~(\ref{eq:bell_diagonal}), which now depends on
the subensemble.  That means, in order to completely specify the state
of all four subensembles, there are 16 real numbers
\(A^{ij},B^{ij},C^{ij}, D^{ij}\) with \(i,j \in\{0,1\}\) required, for
which one obtaines  recurrence relations of the form
\begin{equation}
  \label{eq:recurrence}
  \begin{split}
    A^{(00)}_{n} &\rightarrow
    A^{(00)}_{n+1}(A^{(00)}_n,A^{(01)}_n,\ldots,D^{(11)}_n),\\
    A^{(01)}_{n} &\rightarrow
    A^{(01)}_{n+1}(A^{(00)}_n,A^{(01)}_n,\ldots,D^{(11)}_n),\\
    &\,\,\,\vdots \\ D^{(11)}_{n} &\rightarrow
    D^{(11)}_{n+1}(A^{(00)}_n,A^{(01)}_n,\ldots,D^{(11)}_n).\\
  \end{split}
\end{equation}
These generalize the recurrence relations (\ref{eq:recursion_binary})
for the case of binary pairs, and the relations
(\ref{eq:qpa_recursion}) for the case of noiseless apparatus.

Like the recurrence relations (\ref{eq:qpa_recursion}) and
(\ref{eq:recursion_binary}), respectively, these relations are (modulo
normalization) quadratic forms in the 16 state variables \(\vec {a} =
\left(A^{(00)}, A^{(00)}, \ldots, D^{(11)} \right)^T\), with
coefficients that depend on the error parameters \(f_{\mu\nu}\)
only. In other words, (\ref{eq:recurrence}) can be written in the more
compact form
\begin{equation}
  \label{eq:quadratic_form}
  \vec{a}'_j = \vec{a} M_j \vec{a}^T,
\end{equation}
where, for each \(j \in \{1,\ldots 16\}\), \(M_j\)  is a real \(16\times
16\)-matrix whose coefficients are polynomials in the noise parameters
\(f_{\mu\nu}\).

\subsection{Numerical results}
\label{sec:results_full}
The 16 recurrence relations (\ref{eq:recurrence}) imply a reduced set
of 4 recurrence relations for the quantities
$A_n=\sum_{ij}A_n^{(ij)}$, $\ldots$, $D_n=\sum_{ij}D_n^{(ij)}$ that
describe the evolution of the total ensemble (that is, the
\emph{blend} \cite{englert1999} of the four subensembles) under the
purification protocol. Note that these values are the only ones which
are known and accessible to Alice and Bob, as they have no knowledge
of the values of the error flags. It has been shown in \cite{briegel}
that under the action of the noisy entanglement distillation process,
these quantities converge towards a fixpoint $(A_\infty, B_\infty,
C_\infty, D_\infty)$, where $A_\infty=F_{\text{max}}$ is the maximal
attainable fidelity \cite{duer_briegel}.

\begin{figure}[tp]
  \begin{center}
    \includegraphics[width=.3\figurewidth]{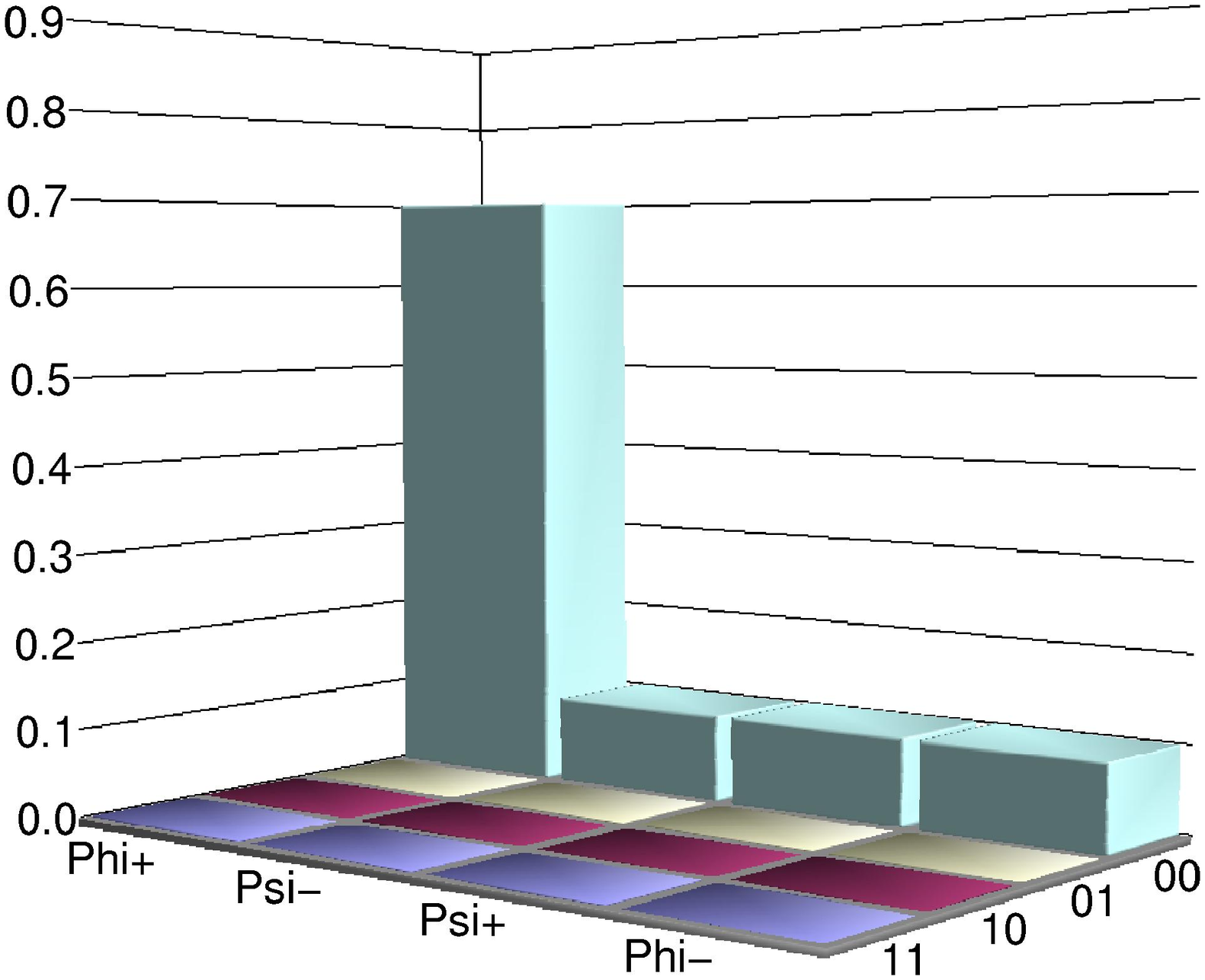}%
    \includegraphics[width=.3\figurewidth]{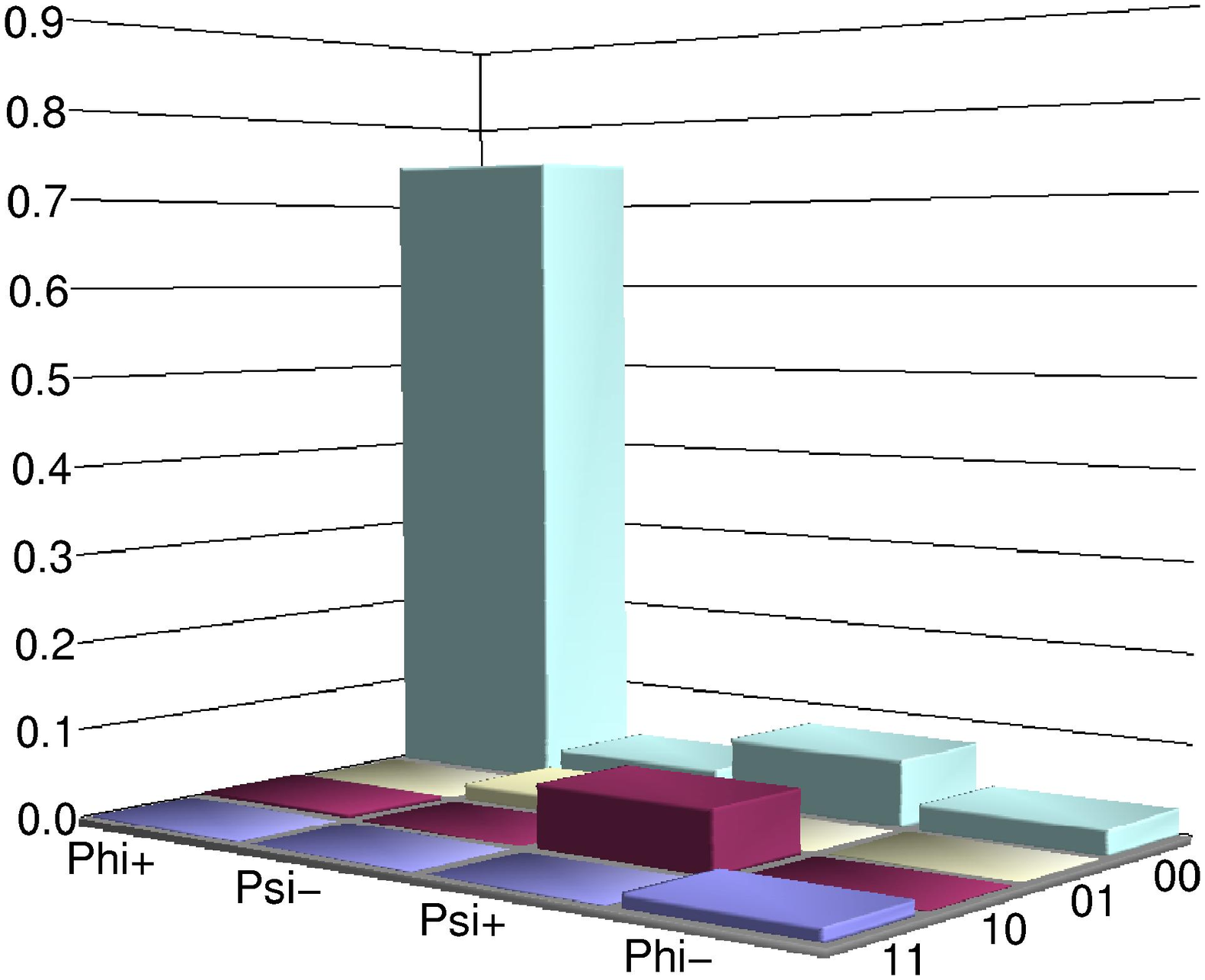}%
    \includegraphics[width=.3\figurewidth]{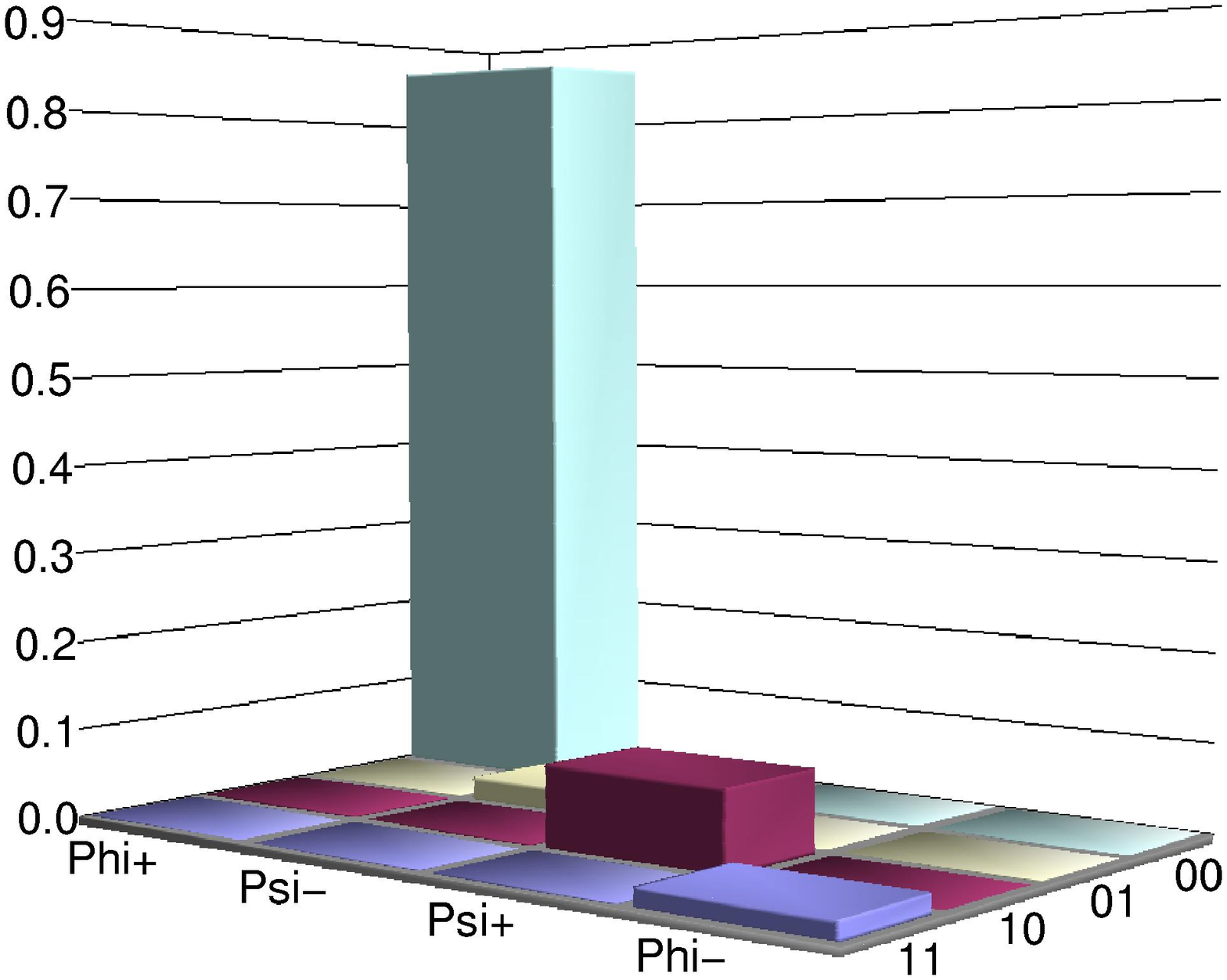}
    \caption{    
      \label{fig:state_evolution}
      Typical evolution of the extended state under the purification
      protocol for the noise parameters $f_{00} = 0.83981, f_{0j} =
      f_{i0} = 0.021131$ and $f_{ij} = 0.003712$ for $i,j \in
      \{1,2,3\}$. This corresponds to a combination of one- and
      two-qubit white noise, as studied in
      \cite{briegel,duer_briegel}, with noise parameters $p_1=0.92$
      and $p_2 = 0.9466$, considering noise in Alice's lab only, or
      $p_1=0.9592$ and $p_2 = 0.973$, considering noise in Alice's and
      Bob's laboratory. We have choosen these values for didactical
      reasons, in order to make the effect more visible.
      }
  \end{center}
\end{figure}

Fig. \ref{fig:state_evolution} shows for typical initial conditions
the evolution of the 16 coefficients \( A_n^{(00)}\ldots
D_n^{(11)}\). They are organized in a $4\times 4$-matrix, where one
direction represents the Bell state of the pair, 
and the other indicates the value of the error flag. The figure shows the state (a)
at the beginning of the entanglement purification procedure, (b) after
few purification steps, and (c) at the fixpoint. As one can see,
initially all error flags are set to zero and the pairs are in a
Werner state with a fidelity of \(70\%\). After a few steps, the
population of the diagonal elements starts to grow; however, none of
the elements vanishes. At the fixpoint, all off-diagonal elements
vanish, which means that there are \emph{strict correlations} between
the states of the pairs and their error flags. 

In order to determine how fast the state converges, we investigate two
important quantities: the first is the fidelity $F_n \equiv A_n$, and
the second is the \emph{conditional fidelity}
\(F_n^{\text{cond}}\equiv A_n^{(00)}+ B_n^{(11)}+ C_n^{(01)}+
D_n^{(10)}\).  Note that the first quantity is the sum over the four
\(\Ket{\Phi^+}\) components in Fig. \ref{fig:state_evolution}, while
the latter is the sum over the four diagonal elements. The conditional
fidelity is the fidelity which Alice and Bob would assign to the pairs
if they knew the values of the error flags, \ie
\begin{equation}
  \label{eq:cond_fidelity}
  F_n^{\text{cond}} = \sum_{i,j} \Bra{\Phi^+} \sigma_{i,j}\rho_{i,j}
  \sigma_{i,j} \Ket{\Phi^+},
\end{equation}
where $\rho_{i,j}$ is the non-normalized state of the subensemble of
the pairs with the error flag $(i,j)$. For convenience, we use the
phase- and spin-flip bits $i$ and $j$ as indices for the Pauli
matrices, \ie \(\sigma_{00} = \mathrm{Id}, \sigma_{01} = \sigma_x,
\sigma_{11} = \sigma_y, \sigma_{10} = \sigma_z\). We will utilize the
advantages of this notation in Section~\ref{sec:flag_update_function}.

\begin{figure}[tp]
  \begin{center}
    \includegraphics[width=0.45\figurewidth]{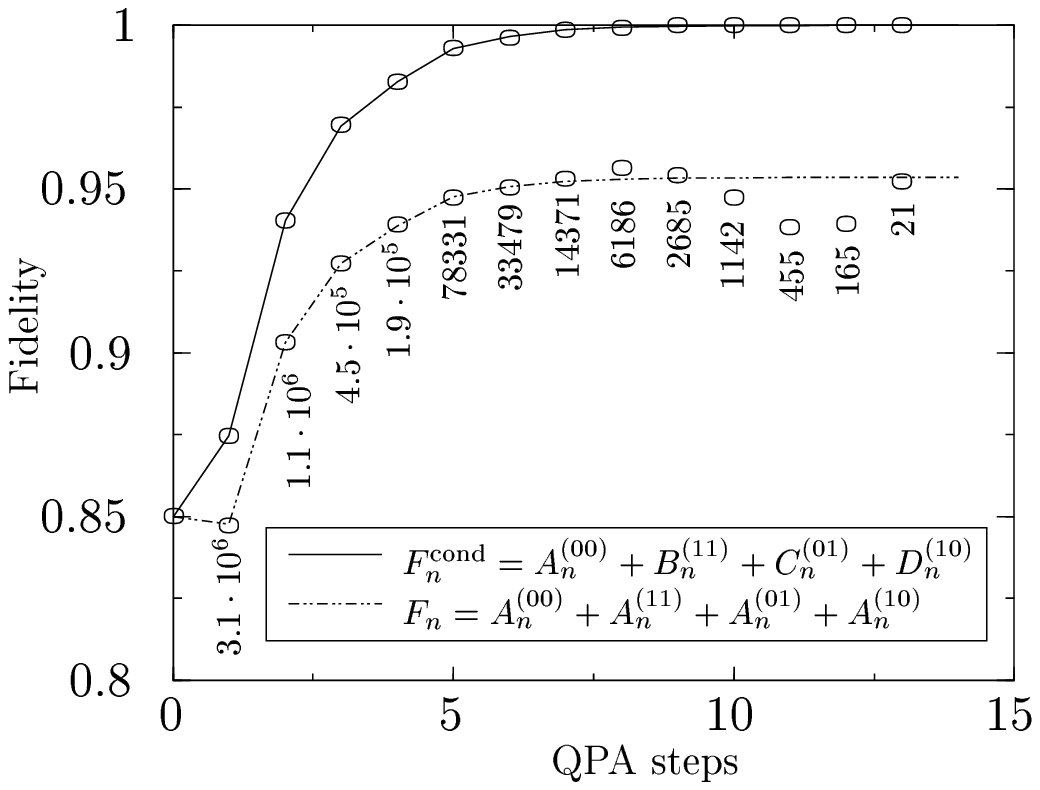}
    \includegraphics[width=0.45\figurewidth]{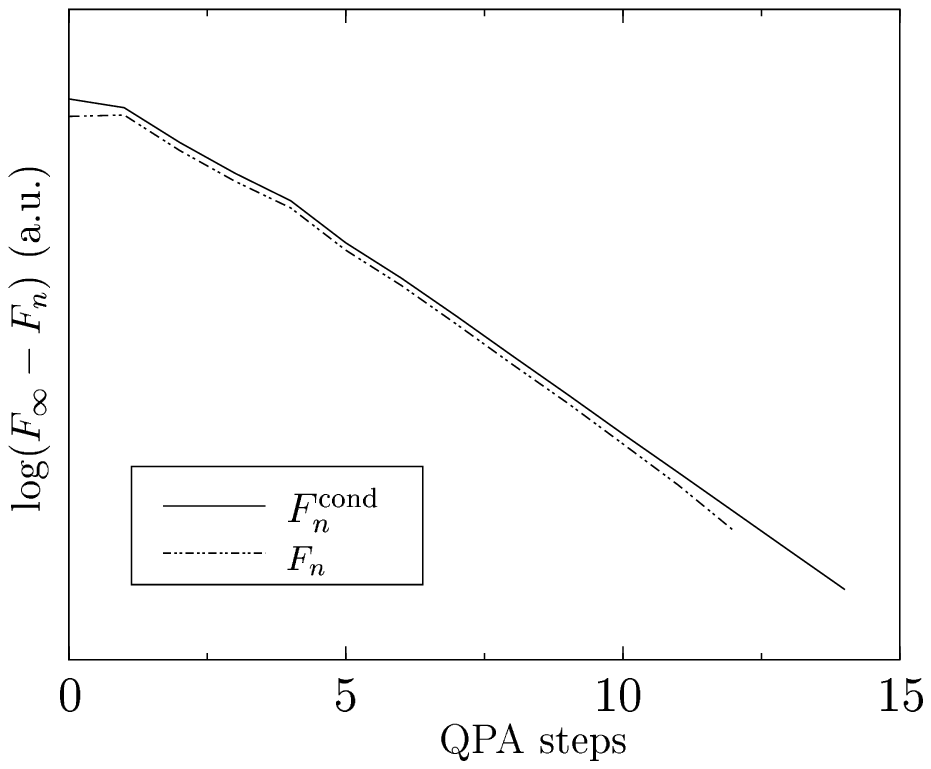}
    \caption{  \label{fig:F_and_F_cond_N}
      The fidelities $F$ and $F_{\text{cond}}$ as a function of
      the number of steps in the security regime of the entanglement
      distillation process (analytical results (lines) and Monte Carlo
      simulation (circles)). The noise parameters for this plot were
      \(f_{00} = 0.91279120\), \(f_{0j} = f_{i0} = 0.0113896\) and
      \(f_{ij} = 0.0020968\) for \(i,j \in \{1,2,3\}\), corresponding
      to white noise with noise parameters \(p_1 = 0.96\) and \(p_2 =
      0.968\) (see Fig.~\ref{fig:state_evolution}). The Monte Carlo
      simulation was started with 10000000 pairs; the numbers indicate
      how many pairs are left after each step of the distillation
      process. This decreasing number is the reason for the increasing
      fluctuations around the analytical curves.}
  \end{center}
\end{figure}

\begin{figure}[htp]
  \begin{center}
    \includegraphics{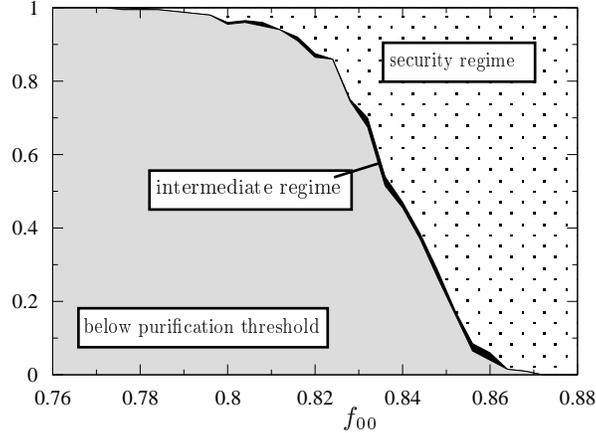}
    \caption{    \label{fig:distri_regimes}
      The size and the location of the three regimes of the
      distillation process. For fixed values of \(f_{00}\), the
      remaining 15 noise parameters \(f_{\mu\nu}\) have been choosen
      at random. Plotted is the relative frequency of finding the
      noise parameters in any of the three regimes as a function of
      \(f_{00}\). } 
  \end{center}
\end{figure}

\begin{figure}[tp]
  \begin{center}
    \includegraphics*[width=\figurewidth]{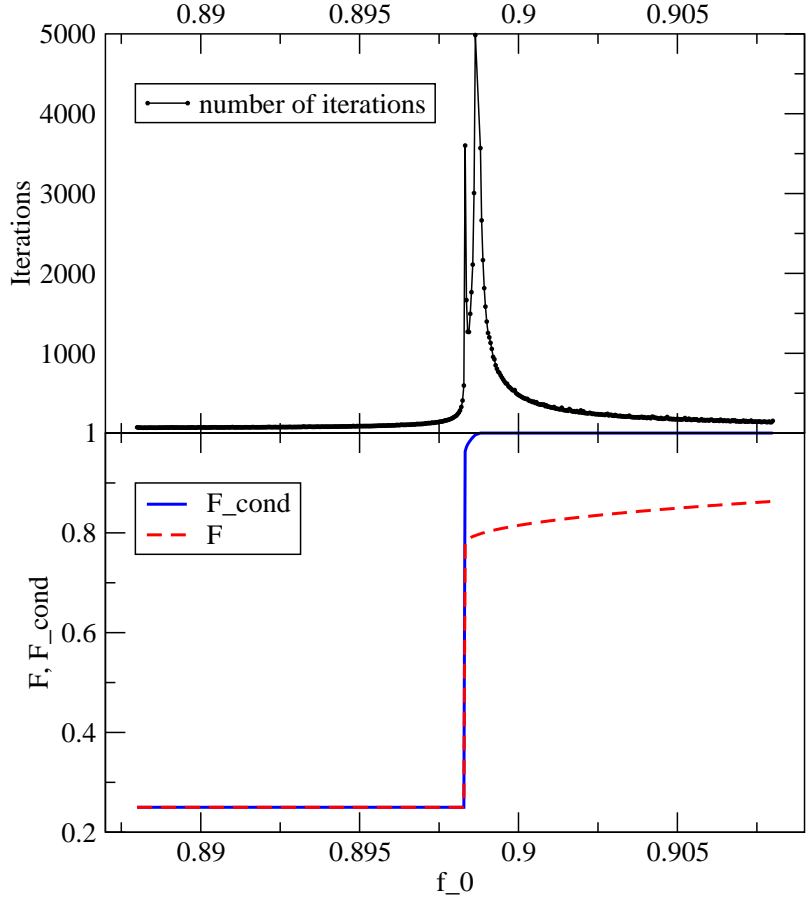}
    \caption{    \label{fig:f_f_cond_iter}
      The effect of one-qubit white noise on the fidelity \(F\),
      the conditional fidelity \(F^{\text{cond}}\) and the number of
      iterations required for the convergence up to an uncertainty
      \(\epsilon = 10^{-12}\).} 
  \end{center}
\end{figure}

The results that we obtain are similar to those for the binary
pairs. We can again distinguish three regimes of noise parameters
\(f_{\mu\nu}\). In the high-noise regime (i.\,e., small values of
\(f_{00}\)), the noise level is above the threshold of the 2--EPP and
both the fidelity \(F\) and the conditional fidelity \(F^\mathrm{cond}\)
converge to the value 0.25. In the low-noise regime (i.\,e.,
large values of \(f_{00}\)), F converges to the maximum fidelity
\(F_\mathrm{max}\) \emph{and} \(F^\mathrm{cond}\) converges to unity
(see Fig. \ref{fig:F_and_F_cond_N}). This regime is the \emph{security
regime}, where we know that secure quantum communication is
possible. Like for binary pairs, there exists also an intermediate
regime, where the 2--EPP purifies but \(F^\mathrm{cond}\) does not
converge to unity.  For an illustration, see
Fig~\ref{fig:distri_regimes}. Note that the size of the intermediate
regime is very small, compared to the security regime. Whether or not
secure quantum communication is possible in this regime is
unknown. However, the answer to this question is irrelevant for all
practical purposes, because in the intermediate regime the
distillation process converges very slowly, as shown in
Fig.~\ref{fig:f_f_cond_iter}. In fact, the divergent behaviour of the
process near the critical points has features remnant of a phase
transition in statistical mechanics.

To estimate the size of the intermediate regime and to compare it to the 
case of binary pairs (Fig.~\ref{fig:fixpoint_map}), we consider the case
of one-qubit white noise, i.e. \(f_{\mu\nu} = f_\mu f_\nu \) and \(f_1
= f_2 = f_3 = (1-f_0)/3\). Here, this regime is known to be bounded by
\[0.8983 < f^\mathrm{crit,lower} < f_0 < f^\mathrm{crit,upper} <
0.8988.\]  
\noindent Note that the size of the intermediate regime is much
smaller than in the case of binary pairs.

\begin{figure}[tp]
  \begin{center}    
    \includegraphics{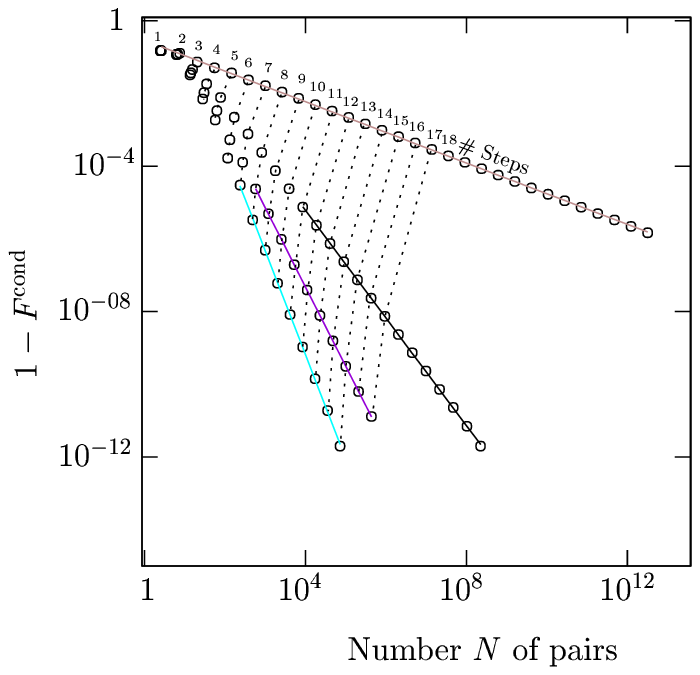}
    \caption{    \label{fig:F-N}
      Number $N$ of pairs needed to create one pair with conditional
      fidelity $F^{\rm cond}$. The initial state of the pairs was of
      the Werner type with fidelity $F_0$ = 85\%. One- and two-qubit
      white noise (see Fig.~\ref{fig:state_evolution}) has
      been assumed with the noise parameters \((p_1, p_2) = (0.9333,
      0.9466)\), \((0.9733,0.9786)\), \((0.9866,0.9833)\),
      \((0.9933,0.9946)\) (from top to bottom).}
  \end{center}
\end{figure}

Regarding the efficiency of the distillation process, it is an
important question how many initial pairs are needed to create one
pair with fidelity $F^{\text{cond}}$, corresponding to the
\emph{security parameter} \(\epsilon \equiv 1 - F^{\text{cond}}\). 
Both the number of required initial pairs
(resources) and the security parameter scale exponentially with the
number of distillation steps, so that we expect a polynomial relation
between the resources and the security parameter
\(\epsilon\). Fig.~\ref{fig:F-N} confirms this relation in a log-log plot
for different noise parameters. The straight lines are fitted
polynomial relations; the fit region is indicated by the lines
themselves.

\subsection{Non-Bell-diagonal pairs}
\label{sec:non_bell_diag}
In the worst-case scenario, Eve generates an ensemble of \(N\) qubit
pairs which she distributes to Alice and Bob. For that reason, Alice
and Bob are not allowed to make specific assumptions on the state of
the pairs.  Most generally, the state of the $2N$ qubits, of which
Alice and Bob obtain $N$ qubits each, can be written in the form
\begin{equation}
\rho_{AB}=\BQuad\sum_{\mu_1\dots\mu_N \atop \mu_1' \dots\mu_N'} \BQuad
\alpha_{\mu_1\CLdots\mu_N \atop \mu_1' \CLdots\mu_N'} |{\cal
B}_{\mu_1}^{(a_1b_1)}\CCdots {\cal B}_{\mu_N}^{(a_Nb_N)}\rangle \!
\langle {\cal B}_{\mu_1'}^{(a_1b_1)}\CCdots {\cal
B}_{\mu_N'}^{(a_Nb_N)}|.
\label{rhoAB_1}
\end{equation}  
Here, $|\mathcal{B}_{\mu_j}^{(a_jb_j)}\rangle$, ${\mu_j}=00,01,10,11$
denote the 4 Bell states associated with the two particles $a_j$ and
$b_j$ and \(j = 1,\ldots,N\).  Specifically, $\Ket{\mathcal{B}_{00}}
\equiv \Ket{\Phi^+} = \left(\Ket{00}+\Ket{11}\right)/\sqrt{2}$,
$\Ket{\mathcal{B}_{01}} \equiv \Ket{\Psi^+} =
\left(\Ket{01}+\Ket{01}\right)/\sqrt{2}$, $\Ket{\mathcal{B}_{10}}
\equiv \Ket{\Phi^-} = \left(\Ket{00}-\Ket{11}\right)/\sqrt{2}$,
$\Ket{\mathcal{B}_{11}} \equiv \Ket{\Psi^-} =
\left(\Ket{01}-\Ket{10}\right)/\sqrt{2}$. In general, (\ref{rhoAB_1})
will be an entangled state of $2N$ particles, which might moreover be
entangled with additional quantum systems in Eve's hands; this allows
for the possibility of so-called coherent attacks
\cite{coll_attacks}. 

Upon reception of all pairs, Alice and Bob apply the following protocol 
to them. Note that steps 1 and 2 are only applied once, while steps 3, 4, 
and 5 are applied recursively in the subsequent distillation process.%
\newline%
{\tt Step 1:} On each pair of particles $(a_j,b_j)$, they apply
randomly one of the four bi-lateral Pauli rotations
$\sigma_k^{(a_j)}\otimes \sigma_k^{(b_j)}$, where k = 0,1,2,3.%
\newline%
{\tt Step 2:} Alice and Bob randomly renumber the pairs,
\((a_j,b_j)\to (a_{\pi(j)},b_{\pi(j)})\) where $\pi(j)$, $j=1,\dots,
N$ is a random permutation.

Steps 1 and 2 are required in order to treat correlated pairs
correctly. Note that steps 1 and 2 would also be required --- as
``preprocessing'' steps --- for the ideal distillation process
\cite{deutsch96}, if one requires that the process converges for
arbitrary states of the form (\ref{rhoAB_1}) to an ensemble of pure
EPR states.  While in \cite{deutsch96} it is possible to check whether
or not the process converges to the desired pure state, by measuring
the fidelity of some of the remaining pairs, this is not possible when
imperfect apparatus is used. Since the maximum attainable fidelity
\(F_{\mathrm{max}}\) is smaller than unity, there is no known way to
exclude the possibility that the non-ideal fidelity is due to
correlations between the initial pairs. In both steps Alice and Bob
discard the information which of the rotations and permutations,
respectively, were chosen by their random number generator.  Thus they
deliberately loose some of the information about the ensemble which is
still available to Eve (as she can eavesdrop the classical information
that Alice and Bob exchange to implement step 1).  After step 1, their
knowledge about the state is summarized by the density operator
\begin{equation}
\tilde\rho_{AB}=\BQuad \sum_{\mu_1\dots\mu_N}\BQuad
p_{\mu_1\dots\mu_N} |{\cal B}_{\mu_1}^{(a_1b_1)}\CCdots {\cal
B}_{\mu_N}^{(a_Nb_N)}\rangle \langle {\cal B}_{\mu_1}^{(a_1b_1)}
\CCdots {\cal B}_{\mu_N}^{(a_Nb_N)}|
\end{equation}    
which corresponds to a \emph{classically correlated ensemble} of pure
Bell states. Since the purification protocol that they are applying in
the following steps maps Bell states onto Bell states, it is
statistically consistent for Alice and Bob to assume after step 1 that
they are dealing with a (numbered) ensemble of pure Bell states, where
they have only limited knowledge about which Bell state a specific
pair is in.  The fact that the pairs are correlated means that the
order in which they appear in the numbered ensemble may have some
pattern, which may have been imposed by Eve or by the channel
itself. By applying step 2, Alice and Bob $(i)$ deliberately ignore
this pattern and $(ii)$ randomize the order in which the pairs are
used in the subsequent purification steps \footnote{This will prevent
  Eve from making use of any possibly pre-arranged order of the pairs,
  which Alice and Bob are ment to follow when performing the
  distillation process.}.  For all statistical predictions made by Alice
and Bob, they may consistently describe the ensemble by the density
operator \footnote{While, strictly speaking, this equality holds only
for $N\to\infty$, the subsequent arguments also hold for the exact but
more complicated form of (\ref{rho_product}) for finite $N$.}
\begin{eqnarray}
\label{rho_product}
\tilde{\tilde\rho}_{AB} &=& \left(\sum_{\mu}p_{\mu} |{\cal
B}_{\mu}\rangle \langle {\cal B}_{\mu}|\right)^{\otimes N} \equiv
(\rho_{ab})^{\otimes N}
\end{eqnarray} 
in which the $p_{\mu}$ describe the probability with which each pair
is found in the Bell state $|{\cal B}_{\mu}\rangle$.  At this point,
Alice and Bob have to make sure that $p_{00}\equiv F> F_{\text{min}}$
for some minimum fidelity $F_{\text{min}}>1/2$, which depends on the
noise level introduced by thier local apparatus. This test can be
performed locally by statistical tests on a certain fraction of the
pairs.

As Alice and Bob now own an ensemble of Bell diagonal pairs, they may
proceed as described in the previous section. However, it is a
reasonable question why Eve cannot take advantage of the additional
information which she has about the state of the pairs: as she is
allowed to keep the information about the twirl operations in step 1
and 2, from her point of view all the pairs remain in an highly
entangled \(2N\)-qubit state. Nevertheless, all predicions made by Eve
must be statistically consistent with the predictions made by Alice
and Bob (or, for that matter, their lab demon), which means that the 
state calculated by Eve must be the same as the state calculated by the 
lab demon, tracing out Eve's additional information. As the lab
demon gets a pure state at the end of the entanglement distillation
process, this must also be the result which Eve obtains using her
additional information, simply due to the fact that no pure state
can be written as a non-trivial convex combination of other states.

\section{How to calculate the flag update function}
\label{sec:flag_update_function}
In this section, we analyse how errors are propagated in the
distillation process. As was mentioned earlier, the state of a given
pair that survives a given purification step in the distillation
process depends on all errors that occured on pairs in earlier steps,
which belong to the ``family tree'' of this pair. Each step of the
distillation process consist of a number of unitary operations followed by a
measurement, which we want to treat separately.

\subsection{Unitary transformations and errors}
\label{sec:unitary_and_errors}

Consider an error \(U_{\text{err}}\) (i.e.\ a random unitary transformation) 
that is introduced before a unitary transformation \(U\) is performed on a state
\(\Ket{\psi}\). Note that, without loss of generality, it is always
possible to split up a noisy quantum operation close to a unitary
operation $U$ in two parts: first, a noisy operation close to
identity, and afterwards the noiseless unitary operation $U$. For that
reason, it only a matter of interpretation whether we think of a
quantum operation which is accompanied by noise, e.\,g. as described
by a master equation of the Lindblad form, or of the combination of
some noise channel first and the noiseless quantum operation afterwards.

We call a transformation \(U_{\text{corr}}\) an \emph{error
corrector}, if the equation
\begin{equation}
  \label{eq:error_corrector}
  U\Ket{\psi} = U_{\text{corr}} U U_{\text{err}}
    \Ket{\psi} 
\end{equation}
holds for all states \(\Ket{\psi}\). Equation
(\ref{eq:error_corrector}) is is obviously solved by \(
U_{\text{corr}}= U U_{\text{err}}^{-1} U^{-1} \).

We want to calculate the error corrector for the Pauli operators and
the unitary operation \(U_\mathrm{2-EPP}\), which consists of the
bilateral $x$-rotations and the BCNOT operation, as described in
Section~\ref{sec:QRandQPA}.

In what follows, it is important to note that Pauli rotations and all
the unitary operations used in the entanglement purification protocol
map Bell states onto Bell states; it is thus expedient to write the
four Bell states as
\begin{equation}
  \label{eq:bell-states}
  \Ket{\bell{ij}} = \frac{1}{\sqrt{2}}\left( \Ket{0j} + (-1) ^
    i\Ket{1\bar{j}}\right) ,
\end{equation}
using the \emph{phase bit} $i$ and the \emph{amplitude bit} $j$ with
\(i,j \in \{0,1\}\) \cite{bennett96a}, which we have implicitly employed in
(\ref{rhoAB_1}). In this notation, we get (ignoring global
phases): \(\sigma_x\Ket{\bell{i,j}} = \Ket{\bell{i,j\oplus 1}}\),
\(\sigma_y \Ket{\bell{i,j}} = \Ket{\bell{i\oplus 1,j\oplus 1}},\) and
\(\sigma_z \Ket{\bell{i,j}} = \Ket{\bell{i\oplus 1,j}}\), where
\(\sigma\) may act on either side of the pair. The \(\oplus\) symbol
indicates addition modulo 2. Consistent with this notation,
\(\sigma_x\) is referred to as the amplitude flip operator,
\(\sigma_z\) as the phase flip operator, and \(\sigma_y\) as the phase
and amplitude flip operator.

The effect of the bilateral one-qubit rotation in the 2--EPP can be
easily expressed in terms of the phase and amplitude bit,
\begin{equation}
  \label{eq:U_x}
  U_x^A \otimes {U_x^{B}}^{-1} \Ket{\bell{i,j}} = \Ket{\bell{i,j\oplus i}},
\end{equation}
and the same holds for the BCNOT operation:
\begin{equation}
  \label{eq:bcnot}
  \text{BCNOT} \Ket{\bell{i,j}}\Ket{\bell{i',j'}} =
  \Ket{\bell{i\oplus i',j}}\Ket{\bell{i',j \oplus j'}} .
\end{equation}

The effect of the unitary part of the 2--EPP onto two pairs in
the states \(\Ket{\bell{i,j}}\) and \(\Ket{\bell{i',j'}}\) can be written
in the form
\begin{equation}
  \label{eq:qpa_bell_pairs}
  U_{\mathrm{2-EPP}} \Ket{\bell{i,j}} \Ket{\bell{i',j'}} =
  \Ket{\bell{i\oplus i',i\oplus j}} \Ket{\bell{i', i' \oplus j' \oplus
      i \oplus j}},
\end{equation}
where the first and second pair plays the role of the ``source'' and
the ``target'' pair. Instead of (\ref{eq:qpa_bell_pairs}), we will
use an even more economic notation of the form \((i,j) \equiv
\Ket{\bell{i,j}}\). Eq. (\ref{eq:qpa_bell_pairs}) can then be written
as
\begin{equation}
  \label{eq:qpa_bell_bits}
  (i,j)(i',j') \stackrel{\mathrm{2-EPP}}{\longrightarrow} (i\oplus
  i',i\oplus j)(i', i' \oplus j' \oplus i \oplus j).
\end{equation}
It is now straightforward to include the effect of the lab demon,
Eq.~(\ref{eq:noise_model}). Applying Pauli rotations \(\sigma_{pa}\)
and \(\sigma_{p'a'}\) to the pairs before the unitary 2--EPP step
(\(\sigma_{00} = \mathrm{Id}, \sigma_{01} = \sigma_x, \sigma_{11} = 
\sigma_y, \sigma_{10} = \sigma_z\)), we obtain:
\begin{equation}
  \label{eq:qpa_pauli_bits}
  \begin{split}
      (i,j)(i',j') 
      \stackrel{\sigma} {\longrightarrow} & (i\oplus p, j\oplus
      a)(i'\oplus p', j'\oplus a')\\ 
      \stackrel{2-EPP} {\longrightarrow} & (i\oplus i' \oplus p\oplus
      p', i\oplus j \oplus p\oplus a) \\
      & (i'\oplus p', i' \oplus j' \oplus i \oplus j \oplus p'\oplus a'
      \oplus p \oplus a).
  \end{split}
\end{equation}
Comparing Eq.~(\ref{eq:qpa_bell_bits}) and
Eq~(\ref{eq:qpa_pauli_bits}), we find that the error corrector for the
error operation \(\sigma_{p,a} \otimes \sigma_{p',a'}\) is given by
\begin{equation}
  U_\mathrm{corr}=\sigma_{p\oplus p', p\oplus a }
  \otimes \sigma_{p',p'\oplus a' \oplus p \oplus a}, 
\end{equation}
independend from the initial state of the pairs. This is the desired
result. 

\subsection{Measurements and measurement errors}
As the 2--EPP does not only consist of unitary transformations but
also of measurements, it is an important question whether or not
errors can be corrected after parts of the system have been measured,
and how we can deal with measurement errors. It is important to note
that whether a pair is kept or discarded in the 2--EPP depends on the
measurement outcomes. This means that, depending on the level of noise
in the distillation process, different pairs may be distilled, each
with a different ``family tree'' of pairs.  This procedure is
conceptually very different from quantum error correction, in the
following sense: In quantum error correction, it is necessary to
correct for errors before performing a readout measurement on a
logical qubit. Here, the situation is quite different: the lab demon
performs all calculations only for bookkeeping purposes.  \emph{No}
action is taken, and thus \emph{no} error correction is performed,
neither by the lab demon, nor by Alice and Bob.

In the analysis of the noisy entanglement distillation process
\cite{briegel,duer_briegel}, not only noisy unitary operations have
been taken into account, but also noisy measurement apparatus, which
is assumed to yield the correct result with the probability \(\eta\),
and the wrong result with the probalility \(1-\eta\).  Surprisingly,
if only the measurements are noisy (\ie all unitary operations are
perfect), the 2--EPP produces \emph{perfect} EPR pairs, as long as the
noise is moderate (\(\eta > 63.5\%\)).  The reason for this property
lies in the fact that \(F=1\) is a fixpoint of the 2--EPP even with
noisy measurements. For a physical understanding of this fact,
it is useful to note that in the distillation process, while
the fidelity of the pairs increases, it becomes more and more unlikely
that a pair which should have been discarded is kept due to a
measurement error. This means that the increasingly dominant effect of
measurement errors is that pairs which should have been kept are
discarded. However, this does not decrease the fidelity of remaining
pairs, only the efficiency of the protocols is affected.

This fact is essential for our goal to extend the concept of error
correctors to the entire 2--EPP which actually includes measurements:
As was shown in \ref{sec:unitary_and_errors}, noise in the unitary
operations can be accounted for with the help of error correctors,
which can be used to keep track of errors through the entire
distillation process; on the other hand, the measurement in the 2--EPP
may yield wrong results due to noise which occured in an earlier
(unitary) operation. This has, however, the same effect as a
measurement error, of which we have seen that it does not jeopardize 
the entanglement distillation process.

\subsection{The reset rule}

From the preceeding two sections, one can identify a first candidate
for the flag update function. The idea is the following: The error
corrector \(U_\mathrm{corr}\) calculated in
\ref{sec:unitary_and_errors} describes how errors on the phase- and
amplitude bit are propagated by the 2--EPP. For the lab demon, this
means that instead of introducing an error operaton \(U_\mathrm{err} =
\sigma_{p,a} \otimes \sigma_{p',a'}\) \emph{before} the unitary part
of the 2-EPP, he could, without changing anything, introduce the
operation \(U_\mathrm{corr}^{-1} =U_\mathrm{corr} =\sigma_{p\oplus p',
p\oplus a } \otimes \sigma_{p',p'\oplus a' \oplus p \oplus a}\) as an
error operation \emph{afterwards}. 

Let us now assume, as an ansatz motivated by the preceeding section, 
that the measurement which follows the unitary operation \(U_\mathrm{2-EPP}\)
does not compromise the concept of error correctors. The lab demon can
then consider the error corrector as an recursive update rule for
errors on the phase- and amplitude bit, \ie. for the phase- and
amplitude error bits which constitute the error flag. Anyway, he has
to discard the target-pair part of the error corrector, as the target
pair is measured and does no longer take part in the distillation
process. The knowledge of the error flag of a specific pair implies
that the lab demon could undo all errors introduced in the family tree
of this pair. For example, if the error flag has the value \((i,j)\),
the lab demon could apply the Pauli operator \(\sigma_{i,j}\) in order
to undo the effect of all errors he introduced up to that point. 

Now it is clear that the noiseless protocol asymptotically produces
perfect EPR pairs in the state \(\bell{0,0}\). It follows that --- in
the asymptotic limit --- a pair with the error flag \((i,j)\) must be
in the state \(\bell{i,j}\), \ie the error flags and the states of the
pairs are strictly correlated. This means, if the assumption above was
true, then the flag update function would be given by \((p,a)(p',a')
\rightarrow (p \oplus p', p\oplus a)\). However, as we will see, the
assumption made above is not quite true; for that reason why we call 
this update function a \emph{candidate} for the flag update function.

The candidate has already the important property that states with
perfect correlations between the error flags (\ie only the
coefficients \(A_{00}, B_{11}, C_{01},\) and \(D_{10}\) are
non-vanishing) are mapped onto states with perfect correlations.

However, using the candidate flag update function, perfect
correlations between flags and pairs are not built up unless they
exist from the beginning. By following the distillation
process in a Monte Carlo simulation that takes the error flags into
account, the reason for this is easy to identify: The population of
pairs which carry an amplitude error becomes too large. Now, the
amplitude bit (not the amplitude \emph{error} bit!) of a target pair is
responsible for the coincidence of Alices and Bobs measurement
results; if the amplitude bit has the value zero, the measurement
results coincide and the source pair will be kept, otherwise it will
be discarded. If the target pair carries an amplitude error, a
measurement error will occur, and there are two possibilities: either
the source pair will be kept even though it should have been
discarded, or \emph{vice versa}, then the source pair will be
discarded although it should have been kept. Obviously, the latter
case does not destroy the convergence of the entanglement distillation
process (but it does have an impact on its efficiency); as Alice
and Bob do not have any knowledge of the error flags, there is nothing
that can be done in this case, and both pairs are discarded. The first
case is more interesting. It is clear that for pairs with
perfectly correlated error flags this case will not occur (due to the
perfect correlations the amplitude error bit can only have the value
one if the amplitude bit has the value one, which is just the second
case). This means that we have the freedom to modify the error flags
of the remaining pair \emph{without} loosing the property that
perfectly correlated states get mapped onto perfectly correlated
states. It turns out that \emph{setting both the error amplitude bit
and the error phase bit of the remaining pair} to zero yields the
desired behaviour of the flag update function, so that perfect
correlations are being built up.

The amplitude error bit of the target pair is given by \(p' \oplus a'
\oplus p \oplus a\). The flag update function can thus be written as
\begin{equation}
  \label{eq:flag_update_function}
  (p,a)(p',a') \rightarrow 
  \left\{ 
    \begin{array}{r l}
      (p\oplus p', p \oplus a) & \mbox{if } p' \oplus a' \oplus p
      \oplus a = 0\\
      (0,0) & \mbox{otherwise}.
    \end{array}
  \right.
\end{equation}
For convenience, the values of the flag update function are given in
Tab.~\ref{tab:error_flags}.

\section{Discussion}
\label{sec:dicussion}
We have shown in Section \ref{sec:fact_of_eve}, 
that the two-way entanglement
distillation process is able to disentangle any eavesdropper from 
an ensemble of imperfect EPR pairs
distributed between Alice and Bob, even in the presence of
noise, \ie when the pairs can only be purified up to a specific
maximum fidelity \(F_\mathrm{max} < 1\). Alice and Bob may use %
these imperfectly purified pairs
as %
a \emph{secure} quantum communication channel. They are thus able to perform
secure quantum communication, and, as a special case, secure classical
communication (which is in this case equivalent to a key distribution
scheme).

In order to keep the argument transparent, we have considered the
case where noise of the form (\ref{eq:noise_model}) is explicitly
introduced by a fictious lab-demon, who keeps track of all error
operations and performs calculations. However, using a simple
indistinguishability argument (see Section \ref{sec:eff_of_noise}), we
could show that any apparatus with the noise characteristics
(\ref{eq:noise_model}) is equivalent to a situation where noise is 
introduced by the lab demon. This means that the security of the
protocol does not depend on the fact whether or not anybody actually
calculates the flag update function. It is sufficient to just use a noisy
2--EPP, in order to get a secure quantum channel.

For the proof, we had to make several assumptions on the noise that
acts in Alices and Bobs entanglement purification device. One
restriction is that we only considered noise which is of the form
(\ref{eq:noise_model}). However, this restriction is only due to
technical reasons; we conjecture that our results are also true for
most general noise models of the form (\ref{eq:operator_sum}). More generally,
a regularization procedure (c.f. Section \ref{sec:eff_of_noise}) can
be used to \emph{actively} make any noise Bell-diagonal.
We have also implicitly introduced the assumption that the eavesdropper 
has no additional knowledge about the noise process, \ie Eve only knows the
publicly known noise characteristics (\ref{eq:noise_model}) of the
apparatus. This assumption would not be justified, for example, if the
lab demon was bribed by Eve, or if Eve was able to manipulate the
apparatus in Alice's and Bob's laboratories, for example by shining in
light from an optical fiber. This concern is not important from a
principial point of view, as the laboratories of Alice and Bob are
considered secure by assumption. On the other hand, this concern has
to be taken into account in a practical implementation.

\begin{acknowledgments}
We thank C.~H.~Bennett, A.~Ekert, G.~Giedke, N.~L\"utken\-haus,
J.~M\"uller-Quade, R.~Rau{\ss}endorf, A.~Schenzle, Ch.~Simon and
H.~Weinfurter for valuable discussions. This work has been supported
by the Deutsche Forschungsgemeinschaft through the
Schwer\-punkts\-programm ``Quanten\-in\-for\-mations\-ver\-ar\-bei\-tung''.

\end{acknowledgments}


\begin{thebibliography}{38}
\expandafter\ifx\csname natexlab\endcsname\relax\def\natexlab#1{#1}\fi
\expandafter\ifx\csname bibnamefont\endcsname\relax
  \def\bibnamefont#1{#1}\fi
\expandafter\ifx\csname bibfnamefont\endcsname\relax
  \def\bibfnamefont#1{#1}\fi
\expandafter\ifx\csname citenamefont\endcsname\relax
  \def\citenamefont#1{#1}\fi
\expandafter\ifx\csname url\endcsname\relax
  \def\url#1{\texttt{#1}}\fi
\expandafter\ifx\csname urlprefix\endcsname\relax\def\urlprefix{URL }\fi
\providecommand{\bibinfo}[2]{#2}
\providecommand{\eprint}[2][]{\url{#2}}

\bibitem[{\citenamefont{Schumacher}(1996)}]{schumacher_noisy_channel}
\bibinfo{author}{\bibfnamefont{B.}~\bibnamefont{Schumacher}},
  \bibinfo{journal}{Phys.~Rev.~A} \textbf{\bibinfo{volume}{54}},
  \bibinfo{pages}{2614} (\bibinfo{year}{1996}).

\bibitem[{\citenamefont{Bennett and Brassard}(1985)}]{bb84}
\bibinfo{author}{\bibfnamefont{C.~H.} \bibnamefont{Bennett}} \bibnamefont{and}
  \bibinfo{author}{\bibfnamefont{G.}~\bibnamefont{Brassard}}, in
  \emph{\bibinfo{booktitle}{Proceedings of IEEE International Conference on
  Computers, Systems and Signal Processing, Bangalore, India}}
  (\bibinfo{publisher}{IEEE}, \bibinfo{address}{New York},
  \bibinfo{year}{1985}), pp. \bibinfo{pages}{175--179}.

\bibitem[{\citenamefont{Ekert}(1991)}]{ekert91}
\bibinfo{author}{\bibfnamefont{A.}~\bibnamefont{Ekert}},
  \bibinfo{journal}{Phys.~Rev.~Lett.} \textbf{\bibinfo{volume}{67}},
  \bibinfo{pages}{661} (\bibinfo{year}{1991}).

\bibitem[{\citenamefont{Ekert et~al.}(1994)\citenamefont{Ekert, Huttner, Palma,
  and Peres}}]{ekert-huttner:94}
\bibinfo{author}{\bibfnamefont{A.~K.} \bibnamefont{Ekert}},
  \bibinfo{author}{\bibfnamefont{B.}~\bibnamefont{Huttner}},
  \bibinfo{author}{\bibfnamefont{G.~M.} \bibnamefont{Palma}}, \bibnamefont{and}
  \bibinfo{author}{\bibfnamefont{A.}~\bibnamefont{Peres}},
  \bibinfo{journal}{Phys.\ Rev.\ A} \textbf{\bibinfo{volume}{50}},
  \bibinfo{pages}{1047} (\bibinfo{year}{1994}).

\bibitem[{\citenamefont{Fuchs and Peres}(1996)}]{fuchs-peres:96}
\bibinfo{author}{\bibfnamefont{C.~A.} \bibnamefont{Fuchs}} \bibnamefont{and}
  \bibinfo{author}{\bibfnamefont{A.}~\bibnamefont{Peres}},
  \bibinfo{journal}{Phys.\ Rev.\ A} \textbf{\bibinfo{volume}{53}},
  \bibinfo{pages}{2038} (\bibinfo{year}{1996}).

\bibitem[{\citenamefont{L\"utkenhaus}(1996)}]{luetkenhaus:96}
\bibinfo{author}{\bibfnamefont{N.}~\bibnamefont{L\"utkenhaus}},
  \bibinfo{journal}{Phys.\ Rev.\ A} \textbf{\bibinfo{volume}{54}},
  \bibinfo{pages}{97} (\bibinfo{year}{1996}).

\bibitem[{\citenamefont{Fuchs et~al.}(1997)\citenamefont{Fuchs, Gisin,
  Griffiths, Niu, and Peres}}]{fuchs-et-al:97}
\bibinfo{author}{\bibfnamefont{C.~A.} \bibnamefont{Fuchs}},
  \bibinfo{author}{\bibfnamefont{N.}~\bibnamefont{Gisin}},
  \bibinfo{author}{\bibfnamefont{R.~B.} \bibnamefont{Griffiths}},
  \bibinfo{author}{\bibfnamefont{C.-S.} \bibnamefont{Niu}}, \bibnamefont{and}
  \bibinfo{author}{\bibfnamefont{A.}~\bibnamefont{Peres}},
  \bibinfo{journal}{Phys.\ Rev.\ A} \textbf{\bibinfo{volume}{56}},
  \bibinfo{pages}{1163} (\bibinfo{year}{1997}).

\bibitem[{\citenamefont{Calderbank and Shor}(1996)}]{CSS}
\bibinfo{author}{\bibfnamefont{A.~R.} \bibnamefont{Calderbank}}
  \bibnamefont{and} \bibinfo{author}{\bibfnamefont{P.}~\bibnamefont{Shor}},
  \bibinfo{journal}{Phys.~Rev.~A} \textbf{\bibinfo{volume}{54}},
  \bibinfo{pages}{1098} (\bibinfo{year}{1996}).

\bibitem[{\citenamefont{Steane}(1996)}]{steane95}
\bibinfo{author}{\bibfnamefont{A.~M.} \bibnamefont{Steane}},
  \bibinfo{journal}{Phys.~Rev.~Lett.} \textbf{\bibinfo{volume}{77}},
  \bibinfo{pages}{793} (\bibinfo{year}{1996}).

\bibitem[{\citenamefont{Bennett
  et~al.}(1996{\natexlab{a}})\citenamefont{Bennett, Brassard, Popescu,
  Schumacher, Smolin, and Wootters}}]{bennett96}
\bibinfo{author}{\bibfnamefont{C.~H.} \bibnamefont{Bennett}},
  \bibinfo{author}{\bibfnamefont{G.}~\bibnamefont{Brassard}},
  \bibinfo{author}{\bibfnamefont{S.}~\bibnamefont{Popescu}},
  \bibinfo{author}{\bibfnamefont{B.}~\bibnamefont{Schumacher}},
  \bibinfo{author}{\bibfnamefont{J.~A.} \bibnamefont{Smolin}},
  \bibnamefont{and} \bibinfo{author}{\bibfnamefont{W.~K.}
  \bibnamefont{Wootters}}, \bibinfo{journal}{Phys.~Rev.~Lett.}
  \textbf{\bibinfo{volume}{76}}, \bibinfo{pages}{722}
  (\bibinfo{year}{1996}{\natexlab{a}}).

\bibitem[{\citenamefont{Bennett
  et~al.}(1996{\natexlab{b}})\citenamefont{Bennett, DiVincenzo, Smolin, and
  Wootters}}]{bennett96a}
\bibinfo{author}{\bibfnamefont{C.~H.} \bibnamefont{Bennett}},
  \bibinfo{author}{\bibfnamefont{D.~P.} \bibnamefont{DiVincenzo}},
  \bibinfo{author}{\bibfnamefont{J.~A.} \bibnamefont{Smolin}},
  \bibnamefont{and} \bibinfo{author}{\bibfnamefont{W.~K.}
  \bibnamefont{Wootters}}, \bibinfo{journal}{Phys.~Rev.~A}
  \textbf{\bibinfo{volume}{54}}, \bibinfo{pages}{3824}
  (\bibinfo{year}{1996}{\natexlab{b}}).

\bibitem[{\citenamefont{Deutsch et~al.}(1996)\citenamefont{Deutsch, Ekert,
  Jozsa, Macchiavello, Popescu, and Sanpera}}]{deutsch96}
\bibinfo{author}{\bibfnamefont{D.}~\bibnamefont{Deutsch}},
  \bibinfo{author}{\bibfnamefont{A.}~\bibnamefont{Ekert}},
  \bibinfo{author}{\bibfnamefont{R.}~\bibnamefont{Jozsa}},
  \bibinfo{author}{\bibfnamefont{C.}~\bibnamefont{Macchiavello}},
  \bibinfo{author}{\bibfnamefont{S.}~\bibnamefont{Popescu}}, \bibnamefont{and}
  \bibinfo{author}{\bibfnamefont{A.}~\bibnamefont{Sanpera}},
  \bibinfo{journal}{Phys.~Rev.~Lett.} \textbf{\bibinfo{volume}{77}},
  \bibinfo{pages}{2818} (\bibinfo{year}{1996}).

\bibitem[{\citenamefont{Mayers}(1996)}]{mayers}
\bibinfo{author}{\bibfnamefont{D.}~\bibnamefont{Mayers}}, in
  \emph{\bibinfo{booktitle}{Advances in Cryptology\,--\,Proceedings of Crypto
  '96}} (\bibinfo{publisher}{Springer-Verlag}, \bibinfo{address}{New York},
  \bibinfo{year}{1996}), pp. \bibinfo{pages}{343--357}, \bibinfo{note}{see also
  quant-ph/9802025}.

\bibitem[{\citenamefont{Biham et~al.}(2000)\citenamefont{Biham, Boyer, Boykin,
  Mor, and Roychowdhurny}}]{biham}
\bibinfo{author}{\bibfnamefont{E.}~\bibnamefont{Biham}},
  \bibinfo{author}{\bibfnamefont{M.}~\bibnamefont{Boyer}},
  \bibinfo{author}{\bibfnamefont{P.~O.} \bibnamefont{Boykin}},
  \bibinfo{author}{\bibfnamefont{T.}~\bibnamefont{Mor}}, \bibnamefont{and}
  \bibinfo{author}{\bibfnamefont{V.}~\bibnamefont{Roychowdhurny}}, in
  \emph{\bibinfo{booktitle}{Proceedings of the Thirty-Second Annual ACM
  Symposium on Theory of Computing}} (\bibinfo{publisher}{ACM Press},
  \bibinfo{address}{New York}, \bibinfo{year}{2000}), pp.
  \bibinfo{pages}{715--724}, \bibinfo{note}{quant-ph/9912053}.

\bibitem[{\citenamefont{Inamori}()}]{hitoshi}
\bibinfo{author}{\bibfnamefont{H.}~\bibnamefont{Inamori}},
  \bibinfo{note}{quant-ph/0008064}.

\bibitem[{\citenamefont{Shor and Preskill}(2000)}]{shor}
\bibinfo{author}{\bibfnamefont{P.~W.} \bibnamefont{Shor}} \bibnamefont{and}
  \bibinfo{author}{\bibfnamefont{J.}~\bibnamefont{Preskill}},
  \bibinfo{journal}{Phys.~Rev.~Lett.} \textbf{\bibinfo{volume}{85}},
  \bibinfo{pages}{441} (\bibinfo{year}{2000}).

\bibitem[{\citenamefont{Bennett et~al.}(1992)\citenamefont{Bennett, Brassard,
  and Mermin}}]{mermin1992}
\bibinfo{author}{\bibfnamefont{C.~H.} \bibnamefont{Bennett}},
  \bibinfo{author}{\bibfnamefont{G.}~\bibnamefont{Brassard}}, \bibnamefont{and}
  \bibinfo{author}{\bibfnamefont{N.~D.} \bibnamefont{Mermin}},
  \bibinfo{journal}{Phys.~Rev.~Lett.} \textbf{\bibinfo{volume}{68}},
  \bibinfo{pages}{557} (\bibinfo{year}{1992}).

\bibitem[{\citenamefont{Briegel et~al.}(1998)\citenamefont{Briegel, D\"ur,
  Cirac, and Zoller}}]{briegel}
\bibinfo{author}{\bibfnamefont{H.-J.} \bibnamefont{Briegel}},
  \bibinfo{author}{\bibfnamefont{W.}~\bibnamefont{D\"ur}},
  \bibinfo{author}{\bibfnamefont{J.~I.} \bibnamefont{Cirac}}, \bibnamefont{and}
  \bibinfo{author}{\bibfnamefont{P.}~\bibnamefont{Zoller}},
  \bibinfo{journal}{Phys.~Rev.~Lett.} \textbf{\bibinfo{volume}{81}},
  \bibinfo{pages}{5932} (\bibinfo{year}{1998}).

\bibitem[{\citenamefont{D\"ur et~al.}(1999)\citenamefont{D\"ur, Briegel, Cirac,
  and Zoller}}]{duer_briegel}
\bibinfo{author}{\bibfnamefont{W.}~\bibnamefont{D\"ur}},
  \bibinfo{author}{\bibfnamefont{H.-J.} \bibnamefont{Briegel}},
  \bibinfo{author}{\bibfnamefont{J.~I.} \bibnamefont{Cirac}}, \bibnamefont{and}
  \bibinfo{author}{\bibfnamefont{P.}~\bibnamefont{Zoller}},
  \bibinfo{journal}{Phys.~Rev.~A} \textbf{\bibinfo{volume}{59}},
  \bibinfo{pages}{169} (\bibinfo{year}{1999}).

\bibitem[{\citenamefont{Giedke et~al.}(1999)\citenamefont{Giedke, Briegel,
  Cirac, and Zoller}}]{giedke}
\bibinfo{author}{\bibfnamefont{G.}~\bibnamefont{Giedke}},
  \bibinfo{author}{\bibfnamefont{H.-J.} \bibnamefont{Briegel}},
  \bibinfo{author}{\bibfnamefont{J.~I.} \bibnamefont{Cirac}}, \bibnamefont{and}
  \bibinfo{author}{\bibfnamefont{P.}~\bibnamefont{Zoller}},
  \bibinfo{journal}{Phys.~Rev.~A} \textbf{\bibinfo{volume}{59}},
  \bibinfo{pages}{2641} (\bibinfo{year}{1999}).

\bibitem[{\citenamefont{Lo and Chau}(1999)}]{lo}
\bibinfo{author}{\bibfnamefont{H.-K.} \bibnamefont{Lo}} \bibnamefont{and}
  \bibinfo{author}{\bibfnamefont{H.~F.} \bibnamefont{Chau}},
  \bibinfo{journal}{Science} \textbf{\bibinfo{volume}{283}},
  \bibinfo{pages}{2050} (\bibinfo{year}{1999}).

\bibitem[{\citenamefont{Macchiavello}(1998)}]{macciavello98}
\bibinfo{author}{\bibfnamefont{C.}~\bibnamefont{Macchiavello}},
  \bibinfo{journal}{Phys.~Lett. A} \textbf{\bibinfo{volume}{246}},
  \bibinfo{pages}{385} (\bibinfo{year}{1998}).

\bibitem[{\citenamefont{Knill et~al.}(1996)\citenamefont{Knill, Laflamme, and
  Zurek}}]{knill}
\bibinfo{author}{\bibfnamefont{E.}~\bibnamefont{Knill}},
  \bibinfo{author}{\bibfnamefont{R.}~\bibnamefont{Laflamme}}, \bibnamefont{and}
  \bibinfo{author}{\bibfnamefont{W.}~\bibnamefont{Zurek}}
  (\bibinfo{year}{1996}), \eprint{quant-ph/9610011}.

\bibitem[{\citenamefont{Preskill}(1998)}]{preskill}
\bibinfo{author}{\bibfnamefont{J.}~\bibnamefont{Preskill}},
  \bibinfo{journal}{Proc.\ Roy.\ Soc.\ Lond.\ A} \textbf{\bibinfo{volume}{454}}
  (\bibinfo{year}{1998}).

\bibitem[{\citenamefont{Knill and Laflamme}(1996)}]{knill1996}
\bibinfo{author}{\bibfnamefont{E.}~\bibnamefont{Knill}} \bibnamefont{and}
  \bibinfo{author}{\bibfnamefont{R.}~\bibnamefont{Laflamme}}
  (\bibinfo{year}{1996}), \eprint{quant-ph/9608012}.

\bibitem[{\citenamefont{Kitaev}(1997)}]{kitaev91}
\bibinfo{author}{\bibfnamefont{A.~Y.} \bibnamefont{Kitaev}},
  \bibinfo{journal}{Russ.~Math.~Surv.} \textbf{\bibinfo{volume}{52}},
  \bibinfo{pages}{1191} (\bibinfo{year}{1997}).

\bibitem[{\citenamefont{Knill et~al.}(1998)\citenamefont{Knill, Laflamme, and
  Zurek}}]{knill1998}
\bibinfo{author}{\bibfnamefont{E.}~\bibnamefont{Knill}},
  \bibinfo{author}{\bibfnamefont{R.}~\bibnamefont{Laflamme}}, \bibnamefont{and}
  \bibinfo{author}{\bibfnamefont{W.}~\bibnamefont{Zurek}},
  \bibinfo{journal}{Science} \textbf{\bibinfo{volume}{279}},
  \bibinfo{pages}{342} (\bibinfo{year}{1998}).

\bibitem[{\citenamefont{Aharonov and Ben-Or}()}]{aharonov96}
\bibinfo{author}{\bibfnamefont{D.}~\bibnamefont{Aharonov}} \bibnamefont{and}
  \bibinfo{author}{\bibfnamefont{M.}~\bibnamefont{Ben-Or}},
  \eprint{quant-ph/9611025}.

\bibitem[{\citenamefont{Bennett et~al.}(1993)\citenamefont{Bennett, Brassard,
  Cr\'epeau, Jozsa, Peres, and Wootters}}]{bennett93}
\bibinfo{author}{\bibfnamefont{C.~H.} \bibnamefont{Bennett}},
  \bibinfo{author}{\bibfnamefont{G.}~\bibnamefont{Brassard}},
  \bibinfo{author}{\bibfnamefont{C.}~\bibnamefont{Cr\'epeau}},
  \bibinfo{author}{\bibfnamefont{R.}~\bibnamefont{Jozsa}},
  \bibinfo{author}{\bibfnamefont{A.}~\bibnamefont{Peres}}, \bibnamefont{and}
  \bibinfo{author}{\bibfnamefont{W.~K.} \bibnamefont{Wootters}},
  \bibinfo{journal}{Phys.~Rev.~Lett.} \textbf{\bibinfo{volume}{70}},
  \bibinfo{pages}{1895} (\bibinfo{year}{1993}).

\bibitem[{\citenamefont{Zukowski et~al.}(1993)\citenamefont{Zukowski,
  Zeilinger, Horne, and Ekert}}]{zukowski1993}
\bibinfo{author}{\bibfnamefont{M.}~\bibnamefont{Zukowski}},
  \bibinfo{author}{\bibfnamefont{A.}~\bibnamefont{Zeilinger}},
  \bibinfo{author}{\bibfnamefont{M.~A.} \bibnamefont{Horne}}, \bibnamefont{and}
  \bibinfo{author}{\bibfnamefont{A.~K.} \bibnamefont{Ekert}},
  \bibinfo{journal}{Phys. Rev. Lett.} \textbf{\bibinfo{volume}{71}},
  \bibinfo{pages}{4287} (\bibinfo{year}{1993}).

\bibitem[{\citenamefont{Pan et~al.}(1998)\citenamefont{Pan, Bouwmeester,
  Weinfurter, and Zeilinger}}]{Pan1998}
\bibinfo{author}{\bibfnamefont{J.-W.} \bibnamefont{Pan}},
  \bibinfo{author}{\bibfnamefont{D.}~\bibnamefont{Bouwmeester}},
  \bibinfo{author}{\bibfnamefont{H.}~\bibnamefont{Weinfurter}},
  \bibnamefont{and}
  \bibinfo{author}{\bibfnamefont{A.}~\bibnamefont{Zeilinger}},
  \bibinfo{journal}{Phys.~Rev.~Lett.} \textbf{\bibinfo{volume}{80}},
  \bibinfo{pages}{3891} (\bibinfo{year}{1998}).

\bibitem[{\citenamefont{van Enk et~al.}(1997)\citenamefont{van Enk, Cirac, and
  Zoller}}]{vanEnk1997}
\bibinfo{author}{\bibfnamefont{S.~J.} \bibnamefont{van Enk}},
  \bibinfo{author}{\bibfnamefont{J.~I.} \bibnamefont{Cirac}}, \bibnamefont{and}
  \bibinfo{author}{\bibfnamefont{P.}~\bibnamefont{Zoller}},
  \bibinfo{journal}{Phys.~Rev.~Lett.} \textbf{\bibinfo{volume}{78}},
  \bibinfo{pages}{4293} (\bibinfo{year}{1997}).

\bibitem[{\citenamefont{van Enk et~al.}(1998)\citenamefont{van Enk, Cirac, and
  Zoller}}]{vanEnk1998}
\bibinfo{author}{\bibfnamefont{S.~J.} \bibnamefont{van Enk}},
  \bibinfo{author}{\bibfnamefont{J.~I.} \bibnamefont{Cirac}}, \bibnamefont{and}
  \bibinfo{author}{\bibfnamefont{P.}~\bibnamefont{Zoller}},
  \bibinfo{journal}{Science} \textbf{\bibinfo{volume}{279}},
  \bibinfo{pages}{205} (\bibinfo{year}{1998}).

\bibitem[{\citenamefont{Briegel et~al.}(2000)\citenamefont{Briegel, D\"ur,
  Cirac, and Zoller}}]{briegel2000}
\bibinfo{author}{\bibfnamefont{H.-J.} \bibnamefont{Briegel}},
  \bibinfo{author}{\bibfnamefont{W.}~\bibnamefont{D\"ur}},
  \bibinfo{author}{\bibfnamefont{J.}~\bibnamefont{Cirac}}, \bibnamefont{and}
  \bibinfo{author}{\bibfnamefont{P.}~\bibnamefont{Zoller}},
  \emph{\bibinfo{title}{The Physics of Quantum Information}}
  (\bibinfo{publisher}{Springer}, \bibinfo{year}{2000}), chap.
  \bibinfo{chapter}{Quantum networks II: Communication over noisy channels}.

\bibitem[{\citenamefont{Kraus}(1983)}]{kraus_states}
\bibinfo{author}{\bibfnamefont{K.}~\bibnamefont{Kraus}},
  \emph{\bibinfo{title}{States, Effects, and Operations}}, vol.
  \bibinfo{volume}{190} of \emph{\bibinfo{series}{Lecture Notes in Physics}}
  (\bibinfo{publisher}{Springer Verlag}, \bibinfo{address}{Berlin Heidelberg
  New York Tokyo}, \bibinfo{year}{1983}).

\bibitem[{\citenamefont{Raussendorf and Briegel}(2001)}]{raussendorf2001}
\bibinfo{author}{\bibfnamefont{R.}~\bibnamefont{Raussendorf}} \bibnamefont{and}
  \bibinfo{author}{\bibfnamefont{H.~J.} \bibnamefont{Briegel}},
  \bibinfo{journal}{Phys.~Rev.~Lett.} \textbf{\bibinfo{volume}{86}},
  \bibinfo{pages}{5188} (\bibinfo{year}{2001}).

\bibitem[{\citenamefont{Englert}(1999)}]{englert1999}
\bibinfo{author}{\bibfnamefont{B.-G.} \bibnamefont{Englert}},
  \bibinfo{journal}{Z. Naturforsch.} \textbf{\bibinfo{volume}{54a}},
  \bibinfo{pages}{11} (\bibinfo{year}{1999}).

\bibitem[{\citenamefont{Cirac and Gisin}(1997)}]{coll_attacks}
\bibinfo{author}{\bibfnamefont{J.~I.} \bibnamefont{Cirac}} \bibnamefont{and}
  \bibinfo{author}{\bibfnamefont{N.}~\bibnamefont{Gisin}},
  \bibinfo{journal}{Phys.~Lett.~A} \textbf{\bibinfo{volume}{229}},
  \bibinfo{pages}{1} (\bibinfo{year}{1997}).

\end{thebibliography}

\end{document}